\documentclass[aps,prb,preprint,amsmath,amssymb,endfloats]{revtex4}
\usepackage[sort&compress]{natbib}
\usepackage{graphicx}
\usepackage{gensymb}
\usepackage{xcolor}
\usepackage{float}

\begin{document}
				
\title{Dissecting Ubiquitin Folding Using the Self-Organized Polymer Model}
\author{Govardhan Reddy}
\email{greddy@sscu.iisc.ernet.in}
 \affiliation{Solid State and Structural Chemistry Unit, Indian Institute of Science, Bangalore, Karnataka, India 560012}
\author{D. Thirumalai} 
\affiliation{Biophysics program, Institute for Physical Sciences and Technology, \\ Department of Chemistry and Biochemistry, University of Maryland, College Park, MD 20742}

\begin{abstract} 
Folding of Ubiquitin (Ub), a functionally important protein found in eukaryotic organisms, is investigated at low and neutral pH at different temperatures using  simulations of the coarse-grained Self-Organized-Polymer model with side chains (SOP-SC).  The melting temperatures ($T_m$s), identified with the peaks in the heat capacity curves, decreases as pH decreases,  in qualitative agreement with  experiments. The calculated radius of gyration, showing dramatic variations with pH, is in excellent agreement with scattering experiments. At $T_m$ Ub folds in a two-state manner at low and neutral pH. Clustering analysis of  the conformations sampled in equilibrium folding trajectories at $T_m$, with multiple transitions between the folded and unfolded states, show a network of metastable states connecting the native and unfolded states.  At low and neutral pH, Ub  folds with high probability through a preferred set of conformations resulting in a pH-dependent dominant folding pathway. Folding kinetics reveal that Ub assembly at low pH occurs by multiple pathways involving a combination of nucleation-collapse and diffusion collision mechanism.  The mechanism by which Ub folds is dictated by the stability of the key secondary structural elements responsible for establishing long range contacts and collapse of Ub. Nucleation collapse mechanism holds if the stability of these elements are marginal, as would be the case at elevated temperatures. If the lifetimes associated with these structured microdomains are on the order of hundreds of $\mu sec$ then Ub folding follows the  diffusion-collision mechanism with intermediates many of which coincide with those found in equilibrium.  Folding at neutral pH is a sequential process with a populated intermediate resembling that sampled at equilibrium. The transition state structures, obtained using a $P_{fold}$ analysis, are homogeneous and globular with most of the secondary and tertiary structures being native-like.  Many of our findings for both the thermodynamics and kinetics of folding are not only in agreement with experiments but also provide missing details not resolvable in standard experiments. The key prediction that folding mechanism varies dramatically with pH is amenable to experimental tests. 
\end{abstract} 

\maketitle

\newpage
 
 \section*{\large{Introduction}}

Major advances in experiments\cite{Schuler08COSB,vzoldak13COSB} and theory \cite{Wolynes95Science,Bryngelson95Proteins,Dill97NSB,Thirumalai99COSB,Shakhnovich06ChemRev,Thirumalai10ARB,Dill12Science}, and creation of coarse-grained models rooted in theory\cite{Hyeon11NatComm,Whitford12RepProg,Tozzini10QRB,Vicatos14Proteins,Best13PNAS} have produced a comprehensive framework for quantitatively describing the way single domain proteins fold.  More recently, technical advances have made it possible to generate long (nearly $milli \ seconds$ in some instances) folding trajectories using atomically detailed simulations in water for several small proteins\cite{Shaw10Science,Lindorff11Science}. These developments have ushered a new era in protein folding in which it is imperative to develop theoretical and computational models so that detailed comparisons with experiments can be made \cite{Thirumalai13COSB}. Many researchers assume that this task requires all atom simulations using empirical force fields. An alternative  is to develop coarse-grained (CG) models, which have proven to have exceptional predictive power not only in the study of self-assembly of proteins but also in understanding larger complexes and biological machines\cite{Hyeon11NatComm,Whitford12RepProg,Vicatos14Proteins}. We use this approach, which we contend is  powerful not to mention computationally tractable, to simulate the folding of Ubiquitin (Ub) at low and neutral pH. 

The importance of the 76-residue Ub, a regulatory protein present in eukaryotic organisms, can hardly be overstated. Ub is involved in a large number of functions ranging from transcriptional regulation to protein degradation and executes these functions by attaching (ubiquitinating) to a number of substrate proteins with great structural diversity. Depending on the function, monoubiquitation \cite{Hicke01NatRevMolCell} and polyubiquitination\cite{Finley09ARBiochem}, both of which are post-translational modifications, have been characterized. In addition to the intrinsic interest in the folding of this small protein, recent studies have established a link between the stability, dynamics, and function of Ub \cite{Lee14JMB}. The monomeric native fold of Ub has five $\beta$-strands, a long $\alpha$-helix, a 3$_{10}$ helix woven together by a complex topology (Fig.~\ref{thermo}A).  The $C_\alpha$ contact-map (Fig.~S1) illustrates prominent interactions between the residues of $\beta_1 \beta_2$ hairpin, and long range contacts involving strands $\beta_1$ and $\beta_5$, and $\beta_3$ and $\beta_5$, and the residues in the loops $L_1$ and $L_2$. Contacts between residues that are distant along the sequence (especially strands $\beta_1$ and $\beta_5$, $\beta_3$ and $\beta_5$, and loops $L_1$ and $L_2$) makes for high contact order, which could be the reason for the complex folding kinetics for a moderate sized protein. The overall folding itself can be accurately estimated using $\tau_F \approx \tau_0 exp(\sqrt(N))$\cite{Thirumalai95JPhysI} where $\tau_0 \approx 1 \mu s$. For the 76-residue Ub, $\tau_F \approx$ 6 ms, which agrees remarkably well with experiments\cite{Khorasanizadeh96NSB,Chung08Biochemistry}.  However, accurate theoretical estimates of $\tau_F$ may be this by itself does not provide insights into the molecular underpinnings of the folding process, which require simulations.

Here, we explore Ub folding using the Self-Organized Polymer model\cite{Hyeon06Structure} with emphasis on native interactions, which has been used to study not only protein folding \cite{Reddy12PNAS} but also several other complex processes ranging from bacterial transcription initiation \cite{Chen10PNAS},  response of microtubule to force \cite{Theisen12JPCB,Kononova14JACS}, and indentation of virus particles \cite{Kononova13BJ}. The model emphasizes native interactions based on the structure of the folded state. The role on non-native interactions, which has been discussed extensively (see below for additional discussions), has been shown to much less dominant in determining the folding of well designed proteins \cite{Camacho95Proteins,Klimov01Proteins,Best13PNAS}. Various aspects of Ub folding have been explored using both atomistic and CG simulations\cite{Fernandez02Physica,Marianayagam04BiophysChem,Alonso95JMB,Alonso98ProteinScience,Irback06Proteins,Sorensen02Proteins,Dastidar05PRE,Kony07ProteinScience,Zhang05Proteins, Piana13PNAS,Mandal14PCCP}. Previous CG simulations, based on $C_{\alpha}$ representation without consideration of electrostatic effects have elucidated the slow dynamics in monomeric Ub at low temperatures \cite{Sorenson02Proteins,Zhang05Proteins} and revealed a change in the folding mechanism as the temperature is lowered \cite{Zhang05Proteins}. Because Ub folding thermodynamics depends dramatically on pH \cite{Wintrode94Proteins} it is crucial to consider electrostatic interactions. Using the SOP model with side chains (SOP-SC) including charge effects, we provide a quantitative description of the thermodynamics and kinetics of folding as a function of pH, which we mimic by modifying the electrostatic interactions. The simulations capture the thermodynamics of Ub folding qualitatively and predict, for the first time, the dimensions in the unfolded state accurately. Interestingly, we predict that although there is a network of connected states linking the folded and unfolded states implying multiple folding pathways, there is a dominant folding path along which Ub self-assembles underscoring the need for probabilistic description of the folding process. The dominant path is found to change dramatically with pH.  Our results for folding thermodynamics and kinetics are in semi-quantitative agreement with a number of experiments, thus establishing that CG models can capture the physics of protein folding.     

\bigskip
 
\section*{\Large{Methods}}

{\bf Self Organized Polymer-Side Chain (SOP-SC) model:} We model the polypeptide chain using the coarse-grained Self Organized Polymer - Side Chain model (SOP-SC)\cite{Hyeon06Structure} in which each residue is represented using two interaction centers, one for the backbone atoms and the other for the side chains (SCs).   Out of the 76 residues in Ub, 23 are charged (Fig.~\ref{thermo}A and S1), which we include in the SOP-SC model (see below). The centers of the interaction centers, representing the backbone atoms and the side chain atoms, are at the $C_\alpha$ atom position of the residue, and the center of mass of the side chain, respectively. The SCs interact via a residue-dependent Betancourt-Thirumalai statistical potential\cite{Betancourt99ProtSci}.  At low pH the acidic residues are protonated, which minimizes the effect of electrostatic interactions on the folding of Ub. To simulate folding at neutral pH we added charges by placing them on the side chains of the charged residues.  The SOP-SC models for Ub are constructed using the crystal structure\cite{Kumar87JMB} (Protein Data Bank (PDB) ID: 1UBQ). 

The energy of a conformation in the SOP-SC models is a sum of bonded (B) and non-bonded (NB) interactions. The interaction between a pair of covalently connected beads (two successive $C_{\alpha}$ atoms or SC connected to a $C_{\alpha}$ atom) is represented by Finite Extensible Nonlinear Elastic (FENE) potential. The non-bonded interactions are a sum of native (N) and non-native (NN) interactions.  If two beads are separated by at least 3 bonds, and if the distance between them in the coarse-grained crystal structure is less than a cutoff distance $R_c$ (Table~S1) then their interactions are considered native. The rest of the pairs of beads, not covalently linked or native, are classified as non-native interactions.

The force-field in the SOP-SC model is:
\begin{equation}\label{pot}
E_{TOT}  = E_{B} + E_{NB}^{N} +  E_{NB}^{NN} + \lambda E^{el}.             
\end{equation} 
The FENE potential, $E_B$, between covalently linked beads is given by 
\begin{equation}\label{fene}
E_{B} = -\sum_{i=1}^{N_{B}}\frac{k}{2}R_{o}^{2} \log \left(1-\frac{(r_{i}-r_{cry,i})^2}{R_{o}^{2}}\right),  
\end{equation}
where $N_{B}$ is the total number of bonds between the beads in the coarse grained model of the polypeptide chain. For Ub, $N_B = 151$. 

The non-bonded native interactions, $E_{NB}^{N}$, in Eq.~\ref{pot} is,
\begin{equation}\label{nat}
\begin{split}
 E_{NB}^{N}& = \sum_{i=1}^{N_N^{bb}} \epsilon_{h}^{bb} \left[ \left(\frac{r_{cry,i}}{r_{i}} \right)^{12} -2 \left(\frac{r_{cry,i}}{r_{i}} \right)^{6} \right] +  \sum_{i=1}^{N_N^{bs}} \epsilon_{h}^{bs} \left[ \left(\frac{r_{cry,i}}{r_{i}} \right)^{12} -2 \left(\frac{r_{cry,i}}{r_{i}} \right)^{6} \right] \\ 
               & +  \sum_{i=1}^{N_N^{ss}} 0.5(\epsilon_{i}^{ss}-0.7) (300.0k_B) \left[ \left(\frac{r_{cry,i}}{r_{i}} \right)^{12} -2 \left(\frac{r_{cry,i}}{r_{i}} \right)^{6} \right]  
\end{split}       
\end{equation}
where $N_{N}^{bb}$(=177), $N_{N}^{bs}$(=486), and $N_{N}^{ss}$(=204) are the numbers of backbone-backbone, backbone-sidechain, sidechain-sidechain native interactions, respectively, $k_B$ is the Boltzmann constant, $r_{i}$ is the distance between the $i^{th}$ pair of residues, and $r_{cry,i}$ is the corresponding distance in the crystal structure. The numbers in the parentheses are for Ub. The strength of interaction between the pair of side chain beads $i$, $\epsilon_{i}^{ss}$, is taken from the Betancourt-Thirumalai statistical  potential\cite{Betancourt99ProtSci}. The values of $\epsilon_{h}^{bb}$ and $\epsilon_{h}^{bs}$ are the same as the ones used in our previous studies on the folding of GFP\cite{Reddy12PNAS}. Thus, the crucial $E_{NB}^{N}$, which determines protein stability, is transferable.

The non-native interactions, $E_{NB}^{NN}$, in Eq.~\ref{pot} is taken to be
\begin{equation}\label{nnat}
\begin{split}
E_{NB}^{NN} & = \sum_{i=1}^{N_{NN}} \epsilon_{l} \left(\frac{\sigma_{i}}{r_{i}} \right)^{6} + \sum_{i=1}^{N_{ang}} \epsilon_{l} \left(\frac{\sigma_{i}}{r_{i}} \right)^{6}
\end{split}
\end{equation}
where  $N_{NN}$(=10159 in Ub) is the total number of non-native interactions, $N_{ang}$(=224 in Ub) is the number of pair of beads separated by 2 bonds in the SOP-SC model, $\sigma_{i}$ is the sum of the radii of the $i^{th}$ pair of residues.   The radii for side chains of amino acids are given in Table~S2. The values of the interaction parameters used in the energy function are given in Table~S1 in the SI. 

Because Ub has 23 charged residues (Fig.~\ref{thermo}A) we expect electrostatic interactions to be important at neutral pH.  The last term in Eq.~\ref{pot} accounts for electrostatic effects, which are modeled using the screened Coulomb potential,
\begin{equation}
E^{el} = \sum_{i=1}^{N_c - 1} \sum_{j=(i +1)}^{N_c} \frac{q_iq_j \ exp(-\kappa r_{ij}) }{\epsilon r_{ij}},
\end{equation}
where $N_c$ is the number of charged residues, $q_i$ and $q_j$ are the charges on the side chains of the $i^{th}$ and $j^{th}$ residues respectively, $\kappa$ is the inverse Debye length, and $r_{ij}$ is the distance between interaction centers located at the centers of mass of side chains $i$ and $j$.  If charges are present on two bonded residues, then electrostatic interactions between these residues is ignored. The value of $q_i$, measured in unit of electron charge, is +1 for positively charged residue and is -1 for negatively charged residues.  In implicit solvent simulations of proteins a range of dielectric constants with values from 2-20 are typically used\cite{Fogolari03Biophys}. We used a value of 10$\epsilon_o$ ($\epsilon_o$  is vacuum permitivity), which gave a reasonable radius of gyration of the protein in the unfolded state. We calculated $\kappa$ assuming a monovalent salt of 10 millimolar is present in the solution. The parameter $\lambda$ in Eq.~\ref{pot} is intended to account for pH effects.  At neutral pH  $\lambda$ = 1.0. In acidic pH, $E^{el}$ is not as relevant and hence we set $\lambda$ = 0. At low pH the charges on the negatively charged residues are neutralized. Because the positively charged residues no longer can engage in salt bridges the polypeptide chain should swell leading to an increase in $R_g$. The residues bearing the positively charged residues would be hydrated, resulting in a reduction in the value of the effective charge ($q_i$). As a result electrostatic static repulsion between the like charges would be softened. Given that these charges are well separated in Ub it stands to reason that interactions between positively charge would not be as strong in the unfolded state as might be naively estimated based on Coulomb's law. So to a first approximation, we neglected the small repulsive interaction, as it is likely to be a perturbation to the hydrophobic interaction. This approximation is not inconsistent with experiments showing that a mutant of S6 in which sixteen charged residues were replaced folded (albeit with altered properties)\cite{Kurnik12PNAS} leading the authors to argue that charge interactions must not be paramount to folding.

The parameters in the SOP-SC energy function are given in the Supplementary Information (SI). 

The rationale for using native-centric CG models to decipher the folding mechanisms of proteins can be traced to several previous computational and theoretical \cite{Wolynes95Science,Bryngelson95Proteins} studies. The earliest studies using lattice models \cite{Camacho95Proteins,Klimov01Proteins}  showed that, for well designed proteins, non-native interactions are likely relevant only during the initial stages of folding resulting in the collapse of the polypeptide chains \cite{Klimov01Proteins}. These findings were also corroborated in certain atomic detailed simulations in implicit solvents \cite{Cardenas03Proteins}. It was also shown that the conformations in the transition state ensemble contain predominantly native interactions \cite{Li00NSB,Klimov01Proteins}, with non-native interactions forming with small probability \cite{Klimov01Proteins}. More recently, analyses based on atomically detailed and $C_\alpha$-Go model simulations for the distribution of the fraction of native contacts in the transition path were found to be similar \cite{Best13PNAS}. These studies reinforce the notion that on time scales exceeding the collapse time it is likely that only native interactions  direct protein folding.   

{\bf Simulations and data analysis:} Following our earlier studies we used\cite{Liu11PNAS,Reddy12PNAS} low friction Langevin dynamics simulations\cite{Veitshans97FoldDes}  to obtain the thermodynamic properties, and Brownian dynamics simulations\cite{Ermak78JChemPhys} to simulate the folding kinetics (see SI for details). 
 
We used the structural overlap function\cite{Guo96JMB} 
$\chi = 1 - \frac{1}{N_{tot}} \sum\limits_{i=1}^{N_{tot}}\Theta\left(\delta-\lvert r_{i}-r_{i}^0\rvert\right)$ to distinguish between different populated states of the protein . Here, $N_{tot} (=11026)$ is the number of pairs of interaction centers in the SOP-SC model of Ub assuming that the interaction centers are separated by at least 2 bonds, $r_{i}$ is the distance between the $i^{th}$ pair of beads, and $r_{i}^0$ being the corresponding distance in the folded state, $\Theta$ is the Heaviside step function, and $\delta = 2$\AA. Examples of folding trajectories at neutral and acidic pH along with the distribution of $\chi$ are displayed in Fig. S2 in the SI. Using $\chi$ as an order parameter,  we calculated the fraction of molecules in the folded and unfolded basins as a function of temperature $T$ (see SI for details).  The radius of gyration $R_g$,  is calculated using $R_g = (1/2N^2)(\sum\limits_{i, j}\vec{r}_{ij}^{ \ 2})^{1/2}$. 

{\bf Identifying the folding network:} In order to determine the network of connected states during Ub folding we clustered the conformations of Ub using a structural metric based on the Distribution of Reciprocal of Interatomic Distances (DRID)\cite{Zhou12JCTC} and leader-like clustering algorithm\cite{Seeber07Bioinfo, Spath80Cluster}. To evaluate the DRID metric, two sets of atoms are required. The first is a set of $n$ centroids, and the second is a set of atoms $N_{atom}$.  The centers of the 74 backbone sites of the SOP-SC model (out of the 76 backbone beads, the 2 termini backbone sites are omitted) are used for both the centroid set ($n=74$), and the distance evaluation set ($N_{atom}=74$).  For each individual centroid $i$, three moments of distribution of reciprocal distances ($\mu_i, \nu_i, \xi_i$) are used to describe the features of atomic distances in a particular conformation. Hence, a  conformation is described by a DRID vector of 222 (=$3 \times n$) components. The geometric distance $\rho$, between two conformations described by the DRID vectors ($\mu_i, \nu_i, \xi_i$) and ($\mu_i^{\prime}, \nu_i^{\prime}, \xi_i^{\prime}$) is obtained by $\rho = \left( (1/3n) \sum\limits_{i=1}^{n} \left[ (\mu_i - \mu_i^{\prime})^2 + (\nu_i - \nu_i^{\prime})^2 + (\xi_i - \xi_i^{\prime})^2 \right] \right)^{1/2}$. The moments of distribution of the reciprocal distances for centroid $i$ ($\mu_i, \nu_i, \xi_i$) are 
\begin{subequations}
\begin{align}
\mu_i = \frac{1}{N_{atom} - 1 - nb_i}\sum^*\limits_{j} \left( 1/d_{ij} \right)  \\
\nu_i = \left[ \frac{1}{N_{atom} - 1 - nb_i}\sum^*\limits_{j} \left( 1/d_{ij} - \mu_i \right)^2 \right]^{1/2} \\
\xi_i = \left[ \frac{1}{N_{atom} - 1 - nb_i}\sum^*\limits_{j} \left( 1/d_{ij} - \mu_i \right)^3 \right]^{1/3},
\end{align}
\label{clustereq}
\end{subequations}
where $nb_i$ is the number of atoms in the distance evaluation set bonded to the centroid $i$, the summation is over the set of atoms assigned for distance evaluation, and the asterisk in Eq.~\ref{clustereq}a indicates that the atoms bonded to the centroid $i$ are omitted in the summation. Two conformations are assigned to different clusters if the geometric distance $\rho$ between them is greater than a certain cutoff value.
 
To identify a suitable cutoff value of $\rho$ for clustering the conformations, the number of clusters $N_{clust}$ as a function of different cutoff values of $\rho$ is calculated (Fig.~S3). The  $N_{clust}$ increases exponentially as the cutoff value for $\rho$ is decreased (Fig.~S3). For the Ub clusters in low pH (Fig.~\ref{cluster}) we use a cutoff value $\rho = 0.0055$\AA$^{-1}$ because for this value 8 clusters with  probabilities greater than 0.01 exist.  Based on the secondary structural content in the 8 clusters, we further coarse-grained them into 5 clusters (Fig.~S4). The cumulative probability of observing a conformation in one  of the 5 clusters exceeds 0.98, which means most of the sampled conformations can be uniquely assigned to one of the major clusters. Similar procedure is used to cluster Ub conformations in neutral pH (see Fig.~\ref{cluster_ele}). The clustering analyses were performed for conformations sampled both at equilibrium and during the folding process.

\bigskip

\section*{\Large{Results and Discussion}}

\begin{figure}
\includegraphics[width=5.8in]{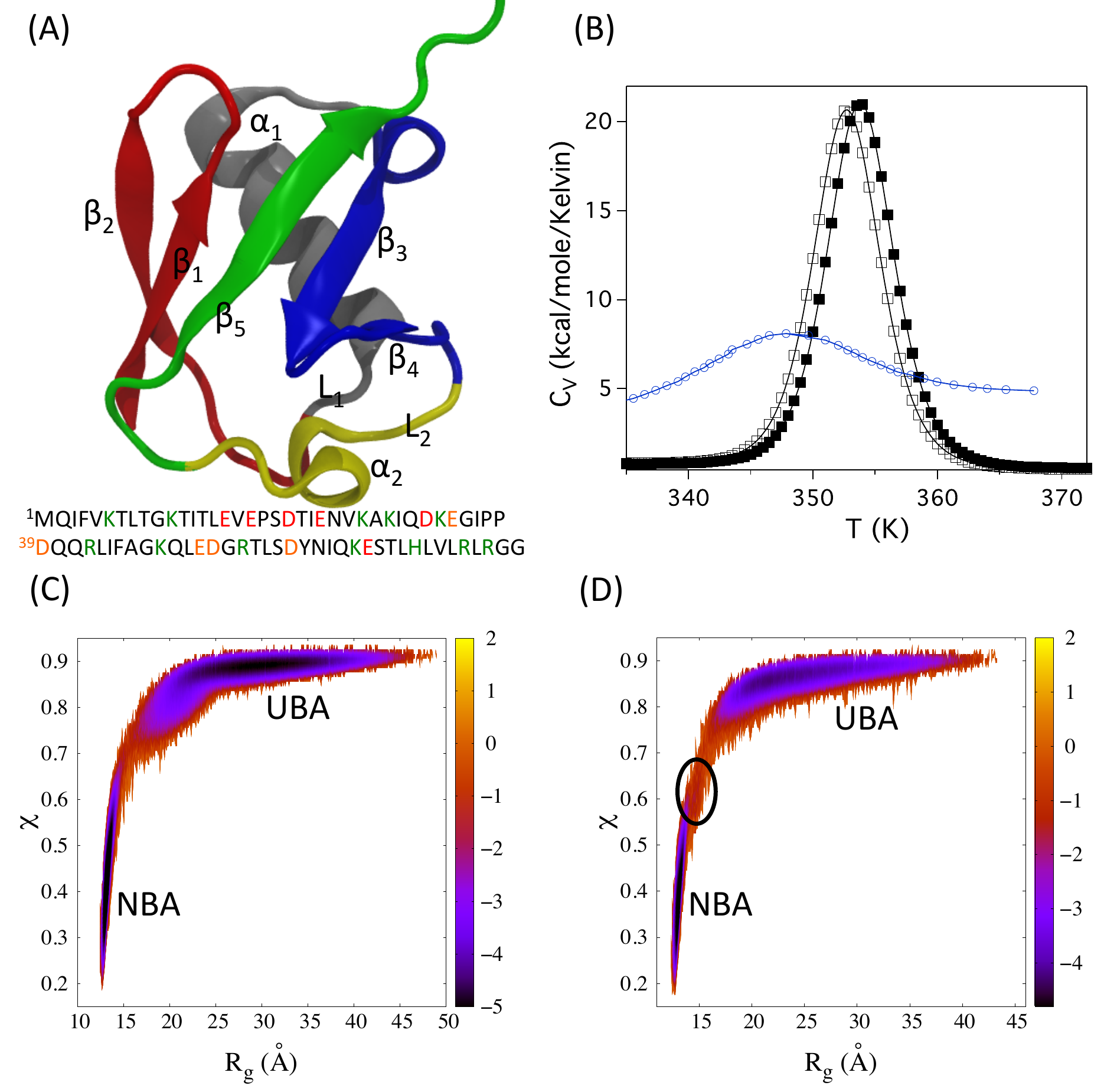}
\caption{Thermodynamics of Ub folding. (a) Ribbon diagram\cite{vmd} of the crystal structure of Ub (PDB ID: 1UBQ).  The strands are labeled $\beta_1$, $\beta_2$, $\beta_3$, $\beta_4$ and $\beta_5$, helix is $\alpha_1$, $3_{10}$ helix is $\alpha_2$, and the interacting loops are $L_1$ and $L_2$. The sequence of Ub is given below. Positively charged residues are in green and negatively charged residues are in red. (B) Heat Capacity $C_v$ as a function of temperature $T$ at low pH (empty squares) and neutral pH (solid squares). For comparison we show the experimental data\cite{Wintrode94Proteins} in blue for heat capacity at pH= 3.0. Although the calculated $T_m$ deviates only by a few degrees from experiments the simulations do not capture the measured peak height and the width of the heat capacity curve. (C) The free energy surface of Ub at low pH and $T_m = 353K$ projected onto the radius of gyration $R_g$, and structural overlap parameter $\chi$, shows 2-state behavior. (D) Same as (C) except the free energy profiles correspond to folding at neutral pH and $T_m = 354K$. The circle highlights a high energy intermediate.}\label{thermo}
\end{figure}

{\bf pH-dependent heat capacity:}   The heat capacity, $C_v(T)$ ($=\frac{\langle E^2 \rangle - \langle E \rangle^2}{k_B T^2}$,  $\langle E(T) \rangle$ and $k_B$ are the average internal energy and Boltzmann constant respectively), as a function of temperature, $T$,  shows that Ub folds in low and neutral pH in an apparent two-state manner (Fig.~\ref{thermo}B). The melting temperature of Ub in low (neutral) pH is $T_m \approx 353K (354K)$. These values are in reasonable agreement  with experiments\cite{Wintrode94Proteins,Molero99Biochemistry}, which show that $T_m$ varies approximately from 320K to $>$ 360K depending on pH. The computed heat heat capacity curves are only in qualitative agreement with calorimetric data\cite{Wintrode94Proteins}. The dramatic changes in the pH-dependent $T_m$ are not quantitatively reproduced. Most importantly, the full width of $C_v(T)$ at half maximum, which changes from $\approx 18K$ at pH = 2 to about $\approx 10K$ in pH = 4 in experiments, is $\approx 5K$ in simulations (Fig.~\ref{thermo}B). The calorimetric enthalpy\cite{Zhou99ProtSci,Liu11PNAS} estimated from the specific heat data in low (neutral) pH is 142.1(142.5) kcal/mole. These values are approximately double the experimental values\cite{Wintrode94Proteins}.   Interestingly, all atom simulations\cite{Piana13PNAS} greatly underestimate the calorimetric enthalpy. In general, it is difficult to compute accurately heat capacity curves using simulations with any empirical force field. In light of this observation, we consider the agreement between simulations and experiments using the same force field (meeting the transferable criterion) as in the our previous reports\cite{Liu11PNAS,Reddy12PNAS} on SH3 and GFP as reasonable.

\begin{figure}
\includegraphics[width=3.3in]{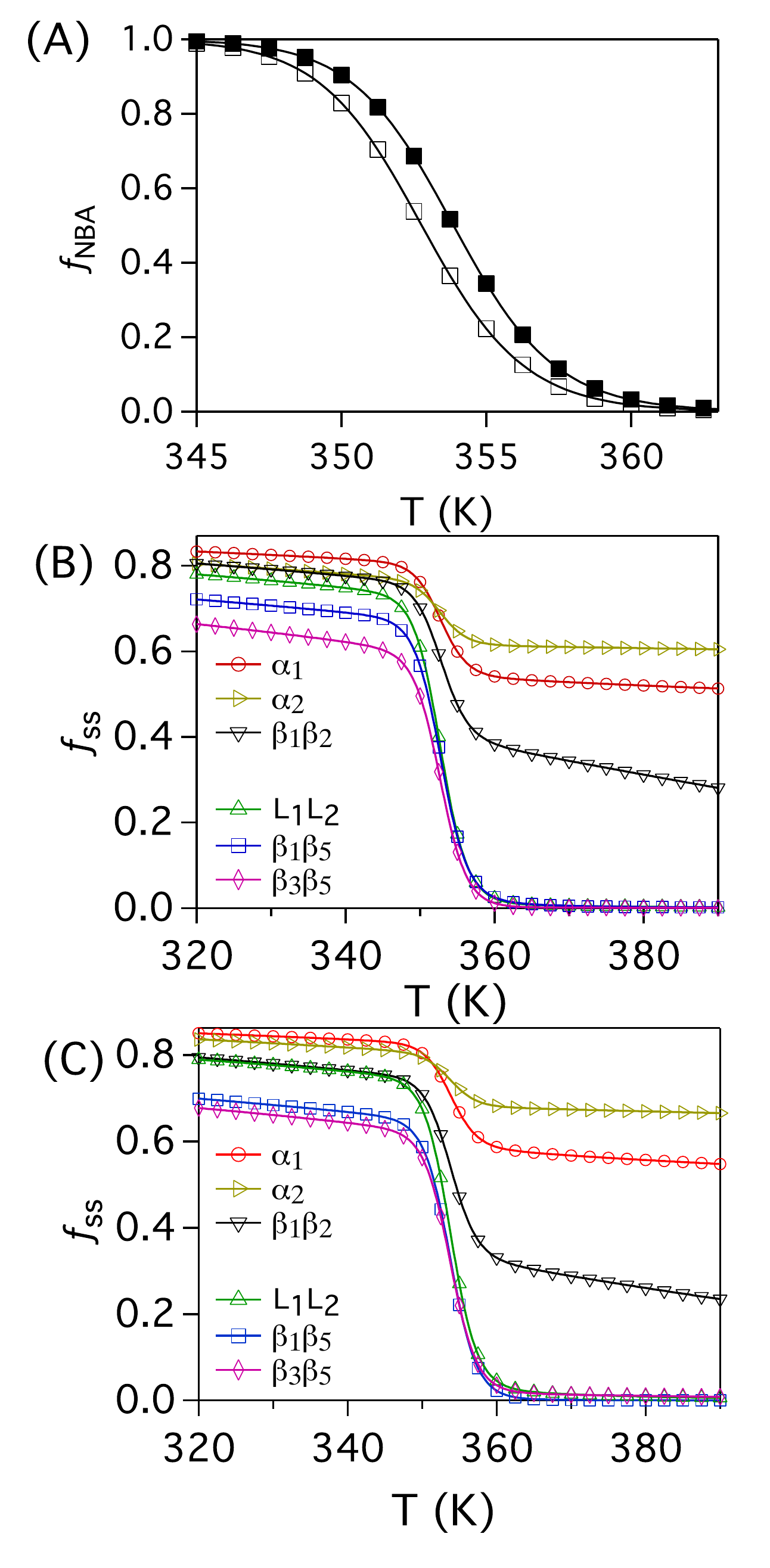}
\caption{(A) Fraction of the protein folded, $f_{NBA}$ in low pH (empty squares) and neutral pH (solid squares) as a function of temperature, $T$. Fraction of the various secondary structural elements, $f_{ss}$ as a function of temperature $T$ in (B) low pH and (C) neutral pH. The secondary structures $\alpha_1$, $\alpha_2$, and $\beta_1 \beta_2$ (Fig.~1A) stabilized by local contacts do not completely unfold even at temperatures above $T_m$. Ub unfolds upon rupture of non-local tertiary interactions involving $\beta_1 \beta_5$, $\beta_3 \beta_5$ and $L_1 L_2$.}\label{frac_secstr_T}
\end{figure}

{\bf Temperature dependent ordering of the NBA and secondary structural elements (SSEs):} 
The temperature dependence of the fraction of Ub in the NBA, $f_{NBA}(T)$, shows a cooperative transition to the folded state (Fig.~\ref{frac_secstr_T}A). The melting temperature, determined using $f_{NBA}(T_m) = 0.5$, coincides with the peak position of $C_v(T)$ (Fig.~\ref{thermo}B). To dissect the ordering of the secondary structural elements, $SSEs$ (= $\beta_1 \beta_2$, $\beta_1 \beta_5$, $\beta_3 \beta_5$, $L_1L_2$, $\alpha_1$, $\alpha_2$), we computed $f_{ss} =\left < N_{ss} \right> / N_{ss}^{o}$ where $\left < N_{ss} \right >$  is the average number of native contacts present in  $SSE$ at $T$, and $N_{ss}^{o}$ is the total number of such contacts in the coarse-grained PDB structure.  The secondary structural elements $\beta_1 \beta_5$, $\beta_3 \beta_5$ and $L_1L_2$, which are stabilized by non-local contacts  are absent in the unfolded Ub. The rupture of these contacts is primarily responsible for the unfolding of the protein at $T > T_m$ in both low and neutral pH (Fig.~\ref{frac_secstr_T}). In contrast,  $\alpha_1$, $\alpha_2$ and $\beta_1 \beta_2$, stabilized by local contacts,  persist even at $T > T_m$ and $\approx 50\%$ of the contacts between the residues stabilizing these structures are present even at $T \approx 400$K  (Fig.~\ref{frac_secstr_T}). Interestingly, snapshots from atomically detailed simulations\cite{Piana13PNAS} at  $T > T_m$ also show (see Fig. 2 in\cite{Piana13PNAS}) persistence of helical structures.

\begin{figure}
\includegraphics[width=3.in]{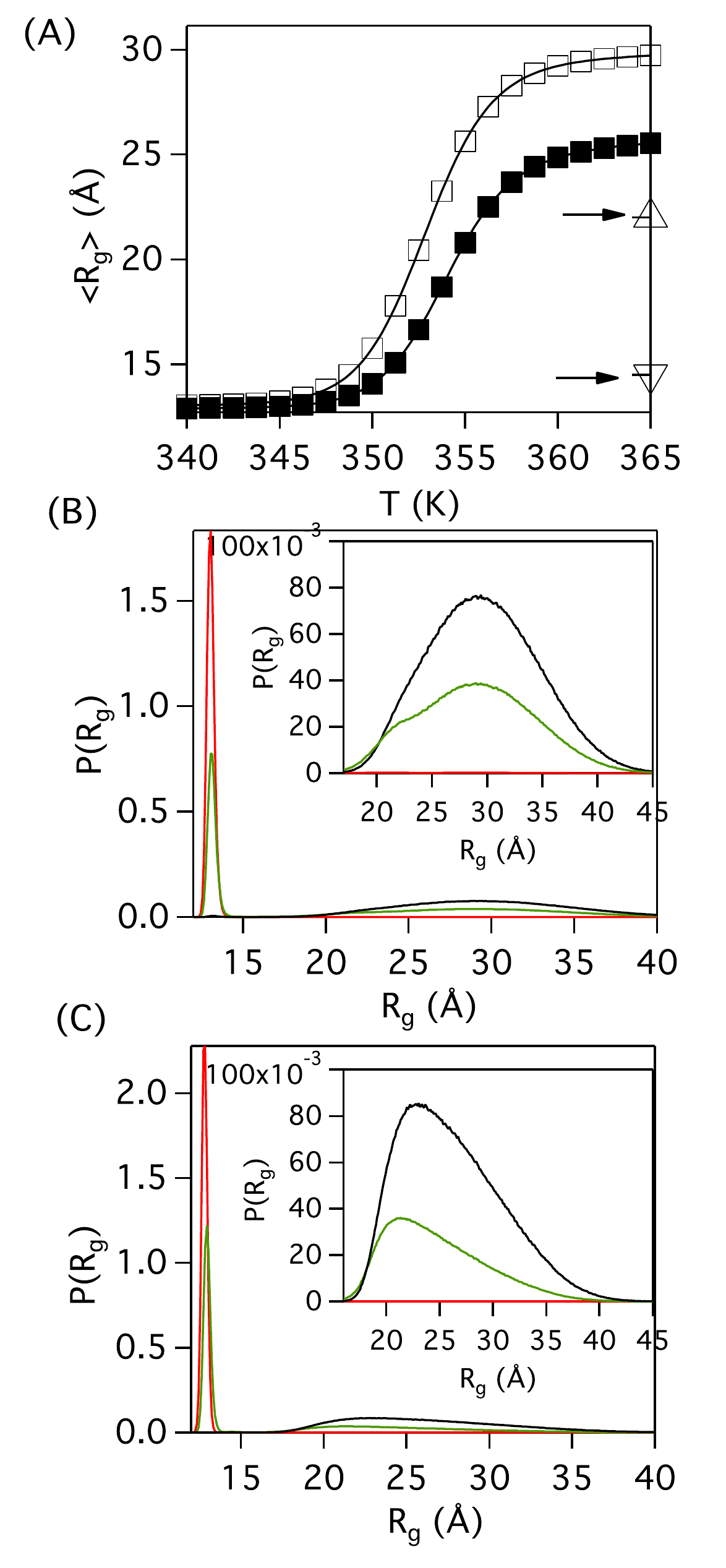}
\caption{(A) $R_g$ as a function of $T$ for low pH (empty squares) and neutral pH (solid squares) calculated from SOP-SC simulations.  Arrows correspond to $\left < R_g \right >$ of Ub in the unfolded state from atomistic simulations\cite{Piana13PNAS} using modified CHARMM22 (black inverted triangle) and OPLS forcefields\cite{Candotti13PNAS} (red triangle) which are $\approx 14.5$\AA \  and $\approx 22$\AA \ respectively. Probability distribution of  $R_g$  (B) low pH at T=343K (red), T=353K(green), T=363K(black) and (C) neutral pH for T=325K (red), T=353K(green), T=375K(black). Inset in (B) and (C) shows $R_g$ distribution in the unfolded state.}\label{Rg_data}
\end{figure}

{\bf pH-dependence of the radius of gyration ($R_g$):}  Plots of average $R_g$ as a function of $T$ show that the dimension of Ub at $T > \approx$ 355K in neutral pH is considerably more compact than in acidic pH (Fig.~\ref{Rg_data}A). The unfolded state $R_g$ distribution at neutral and acidic pH shows conformations with $R_g$ in the range $20\AA \lessapprox R_g \lessapprox 40\AA$ with the peak of the probability distribution at $\approx 30$\AA \ and $\approx 23$\AA \ for high pH and neutral pH respectively (see inset in Fig.~\ref{Rg_data}B and C). The average values of $R_g$ in the unfolded state at high temperatures are $\approx 30$\AA\ and $\approx 25$\AA\ at low and neutral pH, respectively. These values are in excellent agreement with  experiments\cite{Huang12JACS,Gabel09JACS,Jacob04JMB}, which  report a mean $R_g$ of Ub at pH 2.5 and 7.0 are $\approx 32$\AA \ and $\approx 26$\AA \, respectively.  In neutral pH, dominated by attractive electrostatic interactions, Ub samples compact conformations as the centers of mass distance between the secondary structural elements $\beta_1-\beta_5$, $\beta_3-\beta_5$, $L_1-L_2$ are in proximity, thus explaining the compactness (Fig.~S5).   In acidic pH, the interaction between these SSEs are not nearly as strong resulting in expansion of the polypeptide chain (Fig.~S5).

The $\langle R_g \rangle$ of the unfolded state of Ub computed from the probability distribution obtained from atomistic simulations with a modified CHARMM22 potential is $\langle R_g \rangle \approx 14.5$\AA \ (Piana {\it et al.}\cite{Piana13PNAS}, Fig.~S1) compares poorly with the experiments (Fig.~\ref{Rg_data}A). Although Ub  samples conformations with $R_g$ in the range $12\AA \lessapprox R_g \lessapprox 33\AA$ in these simulations\cite{Piana13PNAS}, the peak of the probability distribution is between $12-13$\AA, resulting in $\langle R_g \rangle \approx 14.5$\AA. In contrast, simulations\cite{Candotti13PNAS} using the OPLS force field show that $R_g$ of the unfolded protein is in the range $12\AA \lessapprox R_g \lessapprox 40\AA$, with an estimated $\langle R_g \rangle \approx$ 21-22\AA \ showing that even the sizes unfolded states of  proteins cannot be computed unambiguously using all atom empirical force fields\cite{Piana14COSB}.  More recently, it has been shown that atomic description produces unusually compact unfolded states\cite{Skinner14PNAS}.  Hydrogen exchange experiments\cite{Skinner14PNAS} confirm that the current atomistic forcefields sample compact unfolded conformations with persistent native-like secondary structure due to excessive intramolecular hydrogen bonding. It should be noted that recent computations\cite{Best14JCTC, Piana15JPCB} suggest that by tuning the protein-water interactions\cite{Best14JCTC} or by using a variant of a water model generated by adjusting the dispersion interactions \cite{Piana15JPCB} one can alter the dimensions of the unfolded or intrinsically disordered proteins, providing reasonable  $R_g$ values  in better agreement with the experiments. It remains to be ascertained if these fixes also produce less compact structures for proteins with native states.

{\bf Hint of a high energy intermediate in the free energy surface at neutral pH:} The folding trajectories in Fig.~S2 show that at $T_m$, Ub makes a number of cooperative transitions between the Native Basin of Attraction (NBA) and Unfolded Basin of Attraction (UBA). In acidic  pH,  such transitions between NBA and UBA in a 2-state manner (Fig.~\ref{thermo}C). On the other hand, in neutral pH the NBA $\rightarrow$ UBA involves an intermediate (Fig.~\ref{thermo}D) although its presence is not evident in the specific heat plot (Fig.~\ref{thermo}B). Using these folding trajectories we constructed the free energy surface, $\Delta G(R_g, \chi) = -k_BT_m\ln(P(R_g,\chi))$, where $P(R_g,\chi)$ is the joint probability distribution of $R_g$ and $\chi$ at $T_m$. The $\Delta G(R_g, \chi)$ profiles display two major basins (NBA and UBA) in low and neutral pH conditions (Fig.~\ref{thermo}C and D). These two basins are separated by a barrier (Fig.~\ref{thermo}C and D). In neutral pH the free energy surface has an additional high energy basin, which can be associated with an intermediate (Fig.~\ref{thermo}D).  The shoulder, corresponding to the intermediate, is on the NBA side. Below we show that the shoulder corresponds to the metastable states sampled by Ub in the UBA (see below). 

The $\Delta G(R_g, \chi)$  profiles show that acquisition of the native state occurs only after substantial compaction of the polypeptide chain (Fig.~\ref{thermo}C and D). In both neutral and acidic pH the value of $\chi$, even after a large decrease in $R_g$, is relatively high. We infer that folding only commences after populating an ensemble of minimum energy compact structures\cite{Camacho93PRL}, which has been explicitly demonstrated for Ub folding using single molecule pulling experiments\cite{GarciaManyes09PNAS}. 

\begin{figure}
\includegraphics[width=6in]{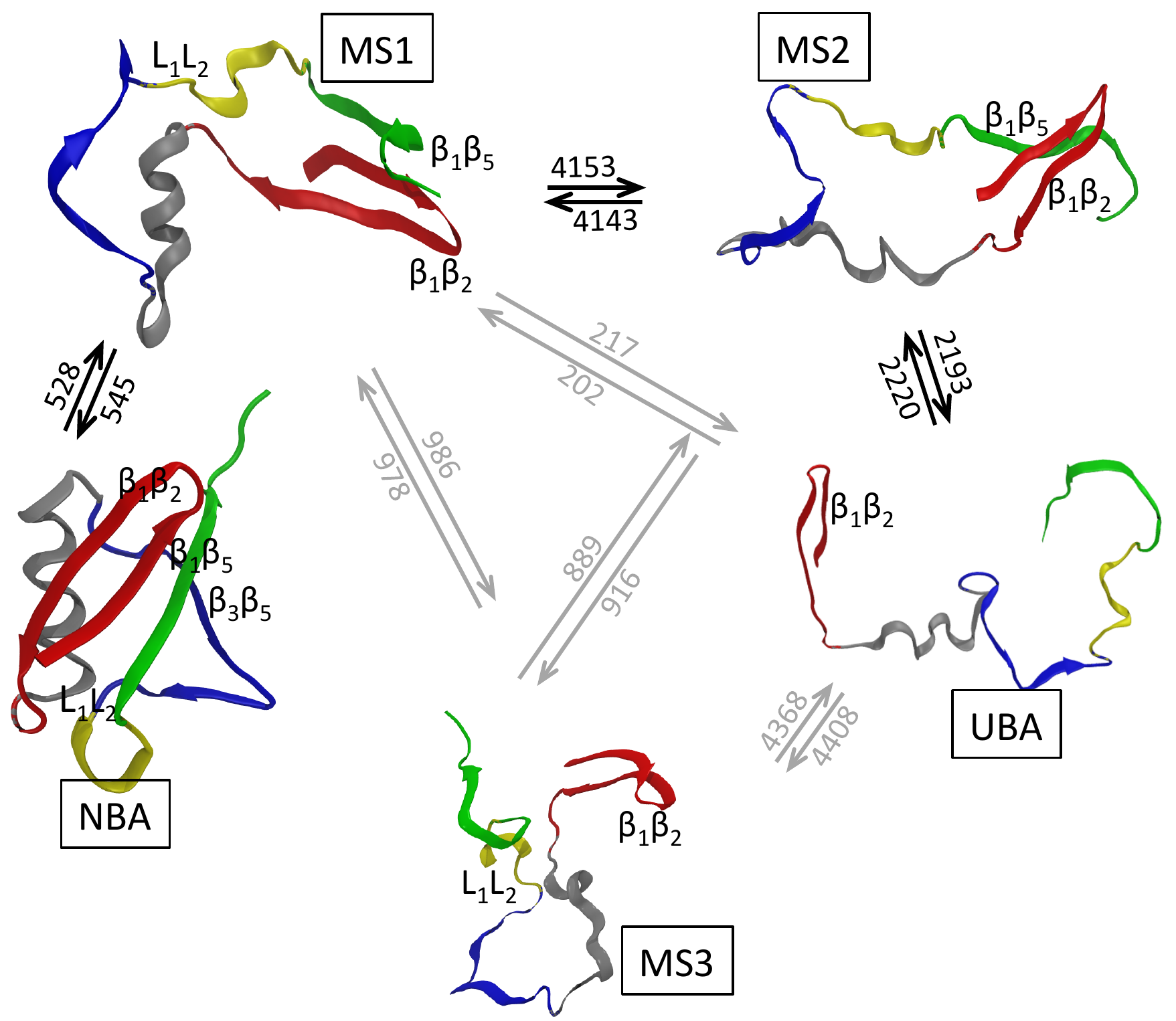}
\caption{Network of connected states obtained using a clustering analysis of the folding trajectories in low pH at $T \approx T_m$ (=353K). There are three metastable clusters labeled MS1-3, which are considered equilibrium intermediates. A representative conformation with the secondary structural elements from each cluster is shown. Dark arrows correspond to the dominant pathway and the grey arrows show possible subdominant routes connecting NBA and UBA. Interestingly, MS3 is not connected to NBA. The numbers show the number of transitions between the clusters in the $37 \mu s$ trajectory.}\label{cluster}
\end{figure}

\begin{figure}
\includegraphics[width=6in]{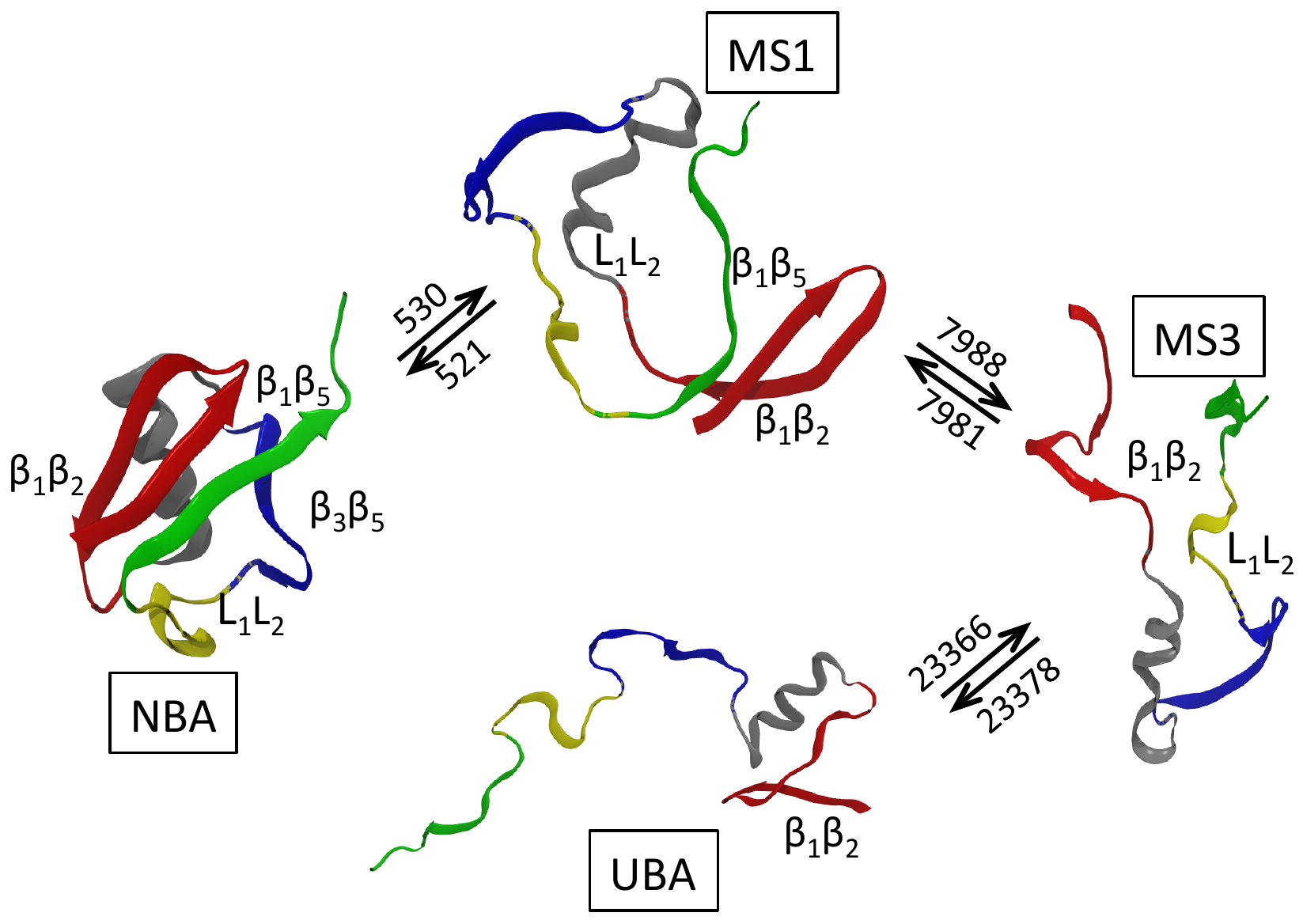}
\caption{Network of connected states obtained using a clustering analysis of the folding trajectories in neutral pH at $T \approx T_m$ (=355K) reveal 4 clusters. In addition to the NBA and UBA, there is a intermediate cluster, MS1 and a metastable cluster MS3. A representative conformation with the secondary structural elements from each cluster is shown. The numbers on the arrows show the number of transitions between the clusters in the $5 \mu s$ trajectory.}\label{cluster_ele}
\end{figure}

{\bf Network of connected states at $T_m$:} In order to assess the complexity of the folding thermodynamics of Ub, we performed a clustering analysis (see Methods) at the melting temperatures using a  $37\mu s$ trajectory at acidic pH and a $5 \mu s$ trajectory in neutral pH (Fig.~S2). Even though at $T_m$, the protein in low and neutral pH appears to fold in a  2-state like manner (Fig.~\ref{thermo}C and D),  the clustering analysis reveals that Ub samples prominent metastable states where it adopts secondary and tertiary structures to varying degrees  (Fig.~\ref{cluster} and \ref{cluster_ele}).

 Of the five prominent clusters identified in acidic pH, three are metastable states labeled  MS1-3 in Fig.~\ref{cluster}. In the MS1 state, the SSEs $\beta_1 \beta_2$, $\beta_1 \beta_5$ and $L_1 L_2$ are formed whereas in MS2 only the  $\beta_1 \beta_2$ hairpin and $\beta_1 \beta_5$ are present.  In the MS3 state the hairpin  $\beta_1 \beta_2$ and $L_1 L_2$ contacts are present. The network of connected states shows that the dominant thermodynamic pathway for assembly of Ub is $ NBA  \leftrightarrows MS1 \leftrightarrows MS2 \leftrightarrows UBA$. Although, conformations belonging to MS1 and MS2 are sampled in the folding trajectories, globally Ub appears as a 2-state folder because the lifetimes of the MS1 and MS2 states are small at $T_m$. (Fig.~\ref{thermo}B, S2). 

In neutral pH, the reversible pathway between UBA and NBA involves  MS1 and MS3 states (Fig.~\ref{cluster_ele}). Electrostatic interactions destabilize the MS2 state, and hence is not sampled  in the folding pathways.    The MS1 state has a long enough lifetime in neutral pH that it is discernible as a  high energy intermediate in the free energy surface in Fig.~\ref{thermo}D. The dominant Ub folding pathway in neutral pH connects the states $ NBA  \leftrightarrows MS1 \leftrightarrows MS3 \leftrightarrows UBA$, which is different from the dominant pathway in low pH. This is the first indication that the folding mechanism depends on pH, which we demonstrate below using kinetic simulations. The MS3, state, which is not a part of the low pH dominant folding pathway (Fig.~\ref{cluster}), is stabilized in neutral pH by favorable interactions among the charged residues at the interface of the SSEs $L_1$, $L_2$ and $\alpha_1$ (Fig.~S1). Interactions associated with these structural elements  play a key role in Ub folding close to neutral pH (see below). 

The thermodynamic pathway can be compared to experiments. Based on experiments\cite{Krantz04JMB, Krantz05JMB} at pH 7.5 and $\Psi$-analysis two folding pathways for Ub were inferred. In the major folding pathway, the hairpin $\beta_1 \beta_2$ forms, then the helix $\alpha_1$ stacks onto $\beta_1 \beta_2$ forming tertiary contacts. Subsequently sheet $\beta_5$ interacts with $\beta_1$ forming the $\beta_1 \beta_5$ contacts. The dominant pathway inferred from experiments on the mutant F45W  agrees partially  with our predictions in neutral pH.  Our analysis (Fig.~\ref{cluster_ele}) reveals that tertiary contacts exist only between $\beta_1 \beta_2$ strands.  We find that the contacts between $\alpha_1$ and $\beta_1 \beta_2$ alone are not stable. We suggest that the next step in the assembly is the formation of contacts between  $\alpha_1$, $L_1$ and $L_2$ due to the charged residues (Fig.~S1), giving rise to transient population of the MS3 state. Finally, sheet $\beta_5$ interacts with $\beta_1$ enabling the formation of the rest of the tertiary contacts. In such compact intermediates where $L_1L_2$ and $\beta_1 \beta_5$ contacts are formed we also observe  interactions between $\alpha_1$ and $\beta_1 \beta_2$ in some of the Ub folding trajectories in neutral pH conditions (see below and structure I3 in Fig.~\ref{ubq_ele_bd}).

The dominant pathway identified using simulations in low pH (Fig.~\ref{cluster}) agrees with the minor pathway inferred from the experiments\cite{Krantz04JMB, Krantz05JMB} at pH 7.5. In this pathway, the hairpin $\beta_1 \beta_2$ forms first, then the sheet $\beta_5$ interacts with $\beta_1$ forming the $\beta_1 \beta_5$ contacts (Fig.~\ref{cluster}, clusters MS1 and MS2). Subsequently, helix $\alpha_1$ forms tertiary contacts with sheets $\beta_1$, $\beta_2$ and $\beta_5$. The UBA cluster is similar to the structure $U1$ in the atomistic simulations\cite{Piana13PNAS} where the contacts between the sheets $\beta_1 \beta_2$ are present. The conformations of Ub in the  MS1 and MS2 states with contacts between $\beta_1 \beta_2$, $\beta_1 \beta_5$ and $L_1L_2$ are similar to the cluster $U1^{\prime}$ in the atomistic simulations. 

\bigskip

{\bf  Temperature and pH dependence of folding mechanism:} 
We generated at least fifty folding Brownian dynamics simulation trajectories in low and neutral pH  at two temperatures to probe the dynamics of Ub folding. The simulations are initiated from an ensemble of unfolded  conformations generated at temperatures $T > T_m$. Regardless of the temperature or pH the SSEs, $\alpha_1$, $\alpha_2$, and $\beta_1 \beta_2$ hairpin are the first structural elements to form. The variations in the self-assembly of Ub occur only in the subsequent stages.

\begin{figure}
\includegraphics[width=6in]{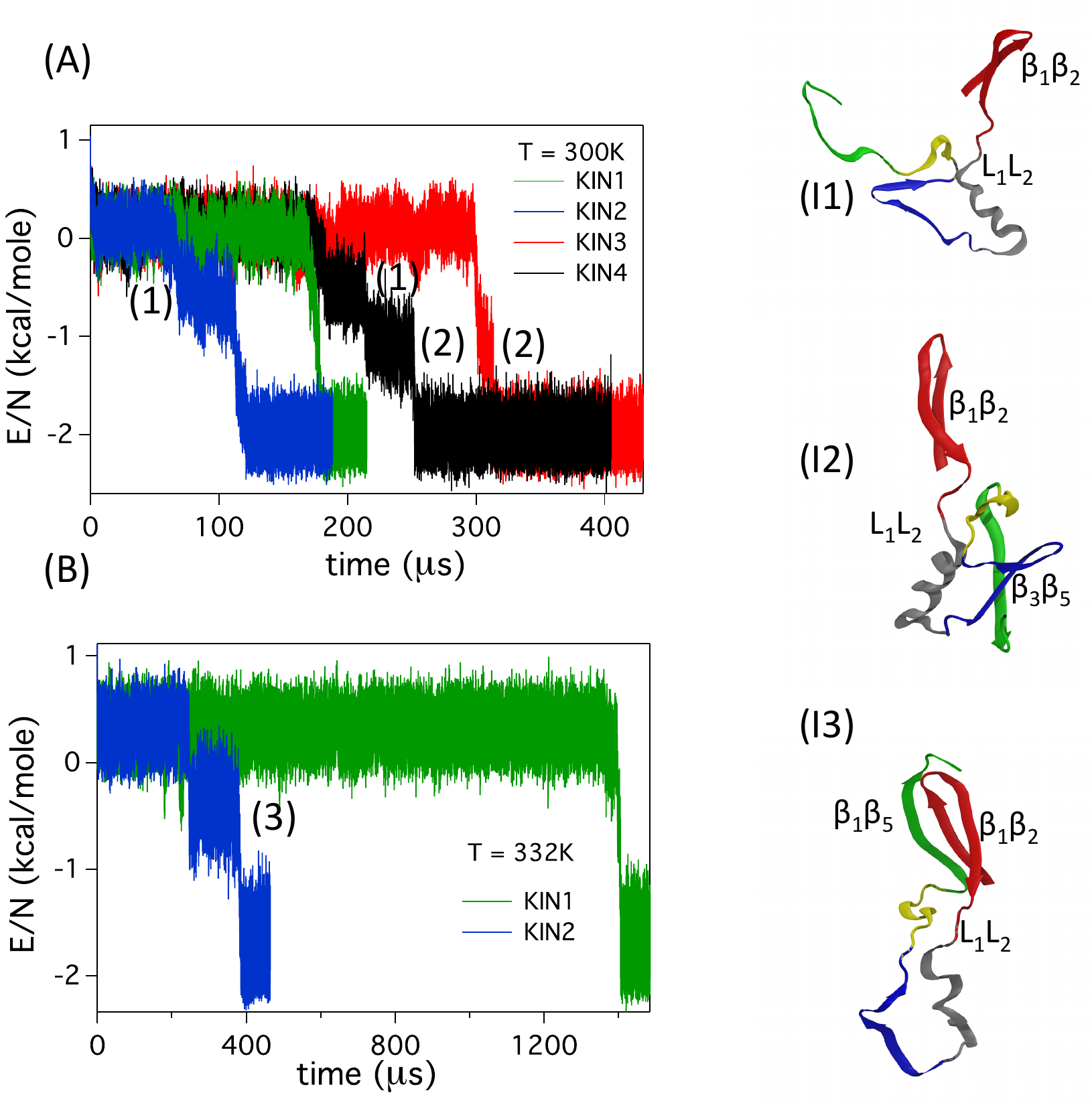}
\caption{Ub folding kinetics in low pH. The folding pathways inferred from change in energy per residue as a function of time, $t$ at (A) $T = 300K$ and (B) $T=332K$. Two kinetic intermediates  I1 and I2 are identified in the folding pathways at T=300K, and the intermediate  I3 is populated  at $T=332K$. Representative structures of the kinetic intermediates are on the right.}\label{kin}
\end{figure}

\begin{figure}
\includegraphics[width=7.25in]{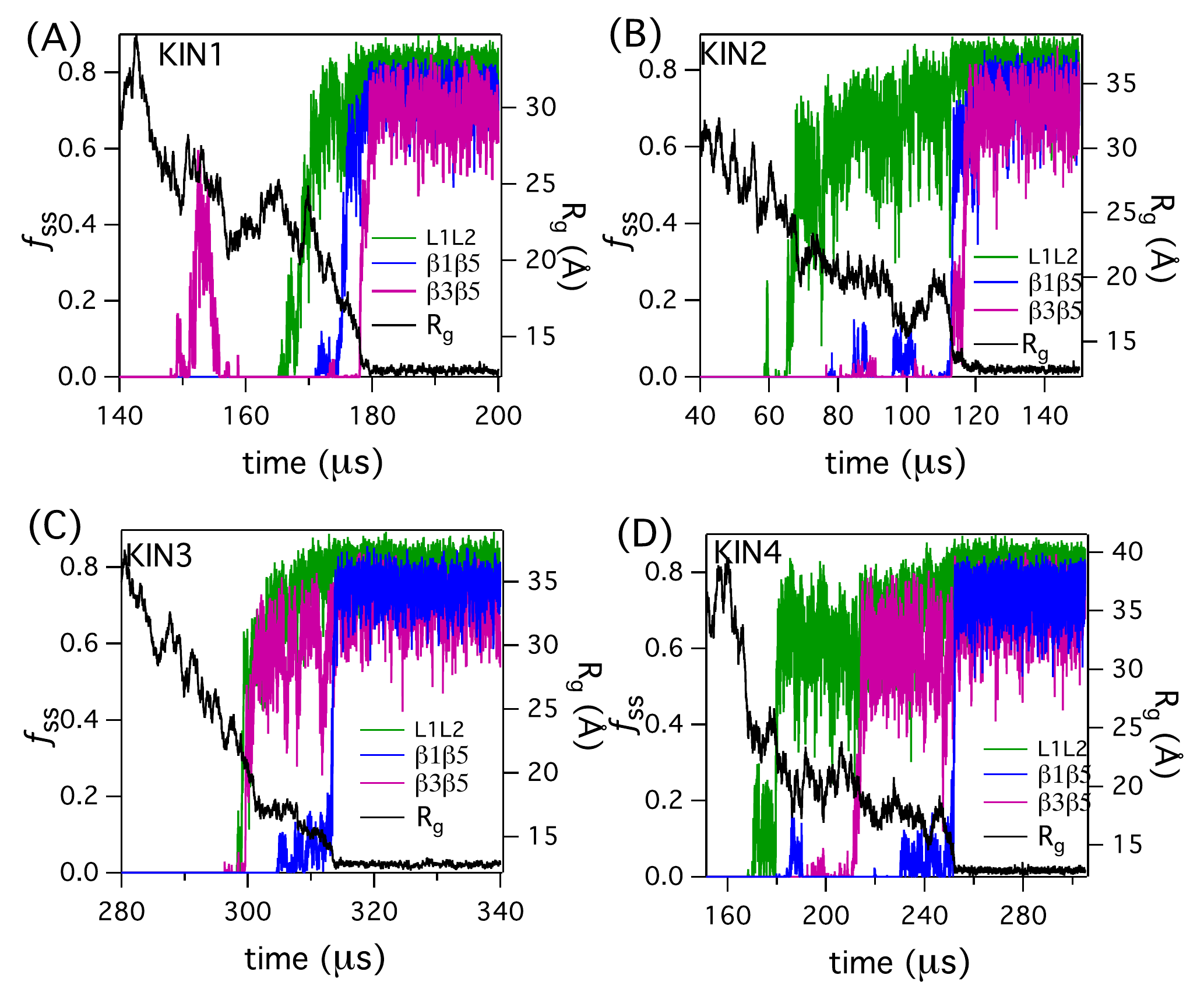}
\caption{Fraction of native contacts in various secondary structural elements ($\beta_1 \beta_5$, $\beta_3 \beta_5$, and $L_1L_2$), $f_{ss}$ and $R_g$ as a function of time in low pH folding trajectories  at $T=300K$. The plots in four panels are for trajectories labeled KIN1, KIN2, KIN3 and KIN4 in Fig.~6A. The green, blue and magenta colors correspond to secondary structures  $L_1L_2$, $\beta_1 \beta_5$, and $\beta_3 \beta_5$, respectively. Radius of gyration, $R_g$ as a function of time is shown in black.} 
\label{kin_T_300K}
\end{figure}

\begin{figure}
\includegraphics[width=6in]{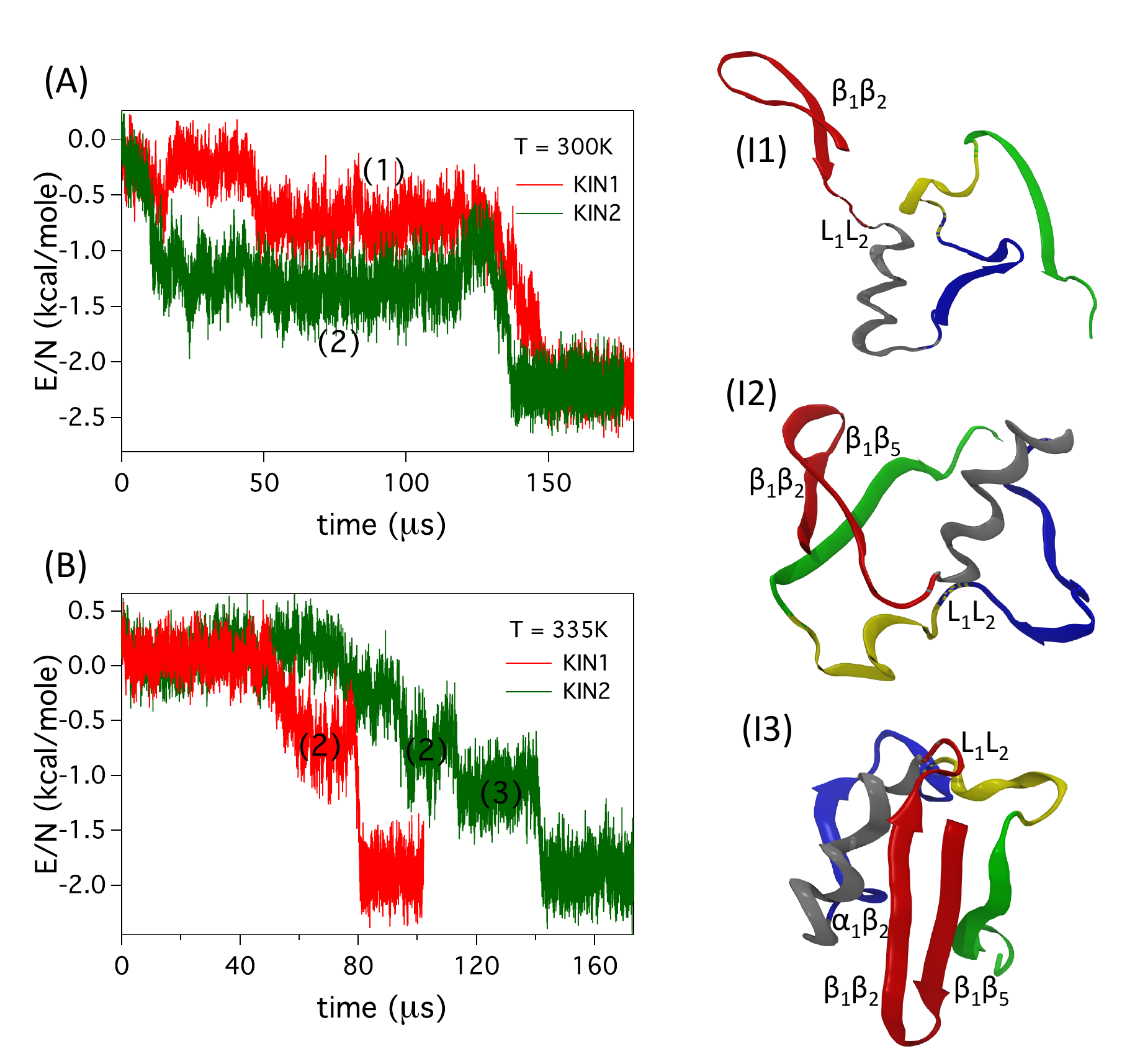}
\caption{Folding kinetics in neutral pH at $T = 300K$ and $335K$. Three intermediates are populated in the folding trajectories and representative structures of the intermediates labeled I1, I2 and I3 are shown on the right.} \label{ubq_ele_bd}
\end{figure}

{\it Low pH:}  At $T= 300K$ Ub folds along four pathways, which we illustrate using one of the folding trajectories (Fig.~\ref{kin}A and \ref{kin_T_300K}). In all the pathways the SSEs $\alpha_1$, $\alpha_2$, and the hairpin $\beta_1\beta_2$ are always present due to their stability. Subsequently there is a bifurcation in the folding pathways. Using energy per residue as a reporter of folding, we find that in the pathways KIN2 and KIN3, one intermediate is populated prior to reaching the folded state whereas there is no persistent intermediate in KIN1 (Fig.~\ref{kin}). The observed bifurcation in the folding pathways, where one route involves a direct UBA $\rightarrow$ NBA transition while in the rest NBA is reached in stages, is the hall mark of the kinetic partitioning mechanism (KPM)\cite{Guo95Biopolymers,Peng08PNAS,Stigler11Science}.   The two intermediates are structurally different (Fig.~\ref{kin}A, I1 and I2). In KIN4, both the intermediates (Fig.~\ref{kin}, I1 and I2) are sampled whereas folding in KIN2, only I1 is sampled and I2 is accessed in KIN3.  I1 is stabilized by the secondary structural elements $\beta_1 \beta_2$ and $L_1 L_2$, whereas I2 is stabilized by the contacts between $\beta_1 \beta_2$, $L_1 L_2$ and $\beta_3 \beta_5$ (Fig.~\ref{kin_T_300K}). There are similarities between I1 and the MS3 state (Fig.~\ref{cluster}) identified in the equilibrium simulations. 

In KIN1, where Ub appears to fold in a 2-state like manner (Fig.~\ref{kin}A), the contacts stabilizing the secondary structures $L_1 L_2$, $\beta_1 \beta_5$ and $\beta_3 \beta_5$ form  almost simultaneously (Fig.~\ref{kin_T_300K}A). They assemble successively separated by a time $\approx 15 \mu s$ (Fig.~\ref{kin_T_300K}A), which is only a fraction of the first passage time. The interactions among the non-local SSEs needed to stabilize the compact states, and collapse of Ub measure by decrease in $R_g(t)$ occur nearly simultaneously (Fig.~\ref{kin_T_300K}A). Folding along this pathway thus follows the nucleation-collapse (NC) mechanism.

In KIN2 and KIN4, the $L_1 L_2$ contacts form ahead of $\beta_1 \beta_5$ and $\beta_3 \beta_5$  leading to I1 (Fig.~\ref{kin_T_300K}B and \ref{kin_T_300K}D). The kinetic intermediate I2 is formed when contacts $L_1 L_2$ and $\beta_3 \beta_5$ are established simultaneously or successively as observed in KIN3 and KIN4 respectively (Fig.~\ref{kin_T_300K}C and \ref{kin_T_300K}D). In the other pathways intermediates with well-defined structures form (Fig.~\ref{kin_T_300K}B and ~\ref{kin_T_300K}C). These figures also show that $R_g(t)$ decrease continues even after some of the non-local SSEs form. The route to the native state in KIN4 is via both the intermediates found in KIN2 and KIN3 (Fig.~\ref{kin_T_300K}D). The presence of a nearly direct transition to the native state in KIN1and folding through the intermediates in the other pathways is in accord with the kinetic partitioning mechanism (KPM). 

The assembly of native Ub along the KIN2, KIN3, and KIN4 pathways at $T=300K$ can be rationalized by the diffusion-collision mechanism (DCM)\cite{Karplus79Biopolymers}. In the nascent stages of folding microdomains (for example $\beta_1 \beta_2$ and $\beta_1 \beta_5$ sheet in I3) form which diffuse freely. Subsequently, some of these collide and coalesce to form the kinetic intermediates with lifetimes on the order of $\approx100\mu s$. In the final stages of folding, the rest of the secondary structural elements collide with the core of the protein structure, and coalesce to form the native structure (see below for further discussion). 

At the higher temperature $T=332K$ Ub folds by the KPM (Fig.~\ref{kin}B). In KIN1, the protein folds in a 2-state manner, and in  KIN2 a single intermediate (Fig.~\ref{kin}B) whose structure  is different from the ones observed at $T=300K$ is populated. This intermediate\cite{Zhang05Proteins, Piana13PNAS}, stabilized by the SSEs $\beta_1 \beta_2$, $L_1L_2$, and $\beta_1 \beta_5$ (Fig.~\ref{kin}, I3),  is similar to the MS1 state at $T_m$, thus providing evidence that Ub samples the equilibrium structures in the process of folding (see below for additional discussion). By analyzing the formation of individual secondary structures (Fig.~S6) we find that at the higher temperature, the contacts which stabilize $L_1L_2$  leading to S1 (MS3 cluster) are unstable (Fig.~S6) in contrast to what is observed at $T=300K$ (Fig.~\ref{kin_T_300K}B and \ref{kin_T_300K}D). The $L_1L_2$ and $\beta_1 \beta_5$ contacts are stabilized simultaneously at $T=332K$ (Fig.~S6) for timescales on the order of $\approx100 \mu s$ leading to I3.  At higher temperatures the contacts between the $\beta$-sheets $\beta_1 \beta_5$ are important in stabilizing the intermediate, as these are the end-to-end contacts in Ub. These interactions minimize the conformational fluctuations  leading to lifetimes that are long enough for S3 to form (Fig.~\ref{kin}B).  At $T=332$K, a 2-state like folding pathway is observed when $\beta_3 \beta_5$ contacts immediately form after structuring of $L_1L_2$ and $\beta_1 \beta_5$ contacts (Fig.~S6A).

{\it Neutral pH:} Folding at neutral pH, where electrostatic interactions play a major role, is dramatically different. In this case, both at $T=335K$ and $300K$ a well populated intermediate is observed (MS1 in Fig.~\ref{cluster_ele}, and I2 in Fig.~\ref{ubq_ele_bd}). In particular, due to the electrostatic interactions between residues in $L_1$ and $L_2$, the $L_1L_2$ loop formation precedes the formation of the I2 intermediate (Fig.~S7). The assembly of Ub occurs by the DCM where preformed micro-domains collide. The intermediate I2 can further become compact due to favorable electrostatic interactions between the helix $\alpha_1$ and  $\beta_2$ strand leading to  I3 (Fig.~\ref{ubq_ele_bd} and S8). This intermediate is not observed in high pH folding. The final stage of compaction is delayed until the micro domains adopt native-like topology (Fig.~S7).  
 
{\bf Comparison with experiments:} It should be borne in mind that there are discrepancies in the interpretations of the different experimental data, which makes it difficult to make direct comparisons with a single experiment. With this caveat, we note that our findings, which are by and large consistent with experiments\cite{Belisle12NSB,Chung08Biochemistry,Rea08Biochemistry,Schanda07PNAS,Belisle07JMB,Crespo06JMB,Larios04JMB,Went04FEBS,Kitahara03PNAS,Qin02JPCB,Cordier02JMB,Khorasanizadeh96NSB,Khorasanizadeh93Biochemistry,Briggs92PNAS}, provide a complete picture of the structures sampled by Ub during folding. However, experiments have only characterized a subset of the predicted intermediates. In accord with the present findings, experiments inferred that Ub folds through an intermediate at low temperatures or mildly denaturing conditions or when mutations  slow down folding.   A common characteristic in all the intermediates, regardless of pH or temperature, is that the $\beta_1 \beta_2$ hairpin is stable for which there is substantial experimental evidence\cite{Cordier02JMB,Schanda07PNAS,Chung08Biochemistry}. The I1 intermediate found here rationalizes experimental studies\cite{Cordier02JMB,Chung05PNAS,Schanda07PNAS}, which provide evidence for a stable $\alpha_1$ helix and unstable $\beta_3$, $\beta_4$ and $\beta_5$ strands. In addition  protein vivisection suggested\cite{Zheng10JMB} an intermediate structure of Ub ubiquitin in which the small  $\beta_4$ (Fig~\ref{thermo}A) strand is unstructured. (Fig.~\ref{kin} and \ref{ubq_ele_bd}B).   Taken together we conclude that our simulations provide a complete structure  of the populated intermediates filling in gaps in experimental studies.

The environment-dependent complex folding pathways are captured in Fig.~\ref{schematic}. The folding mechanisms, involving characterization of the network of connected intermediate structures and transitions between them, are vastly different if the external conditions are altered. The most general characteristic is that folding is a stochastic process in which assembly occurs by multiple pathways. Depending on the conditions the flux through these pathways can be altered and as demonstrated here one can even a single dominant pathway for folding.  Validation of our predictions requires experiments probing folding kinetics as function of pH and temperature.

{\bf Coincidence of equilibrium and kinetic intermediates:} The structures of the I1 kinetic intermediates in low pH at $T=300K$, and I3 at $T=332K$ (Fig.~\ref{kin}) are similar to those observed in the MS1 and MS3 states (Fig.~\ref{cluster}). The dominant folding/unfolding pathway identified at $T_m$  (Fig.~\ref{cluster}) is very similar to the KIN2 folding pathway at $T=335K$. In neutral pH the dominant folding pathway (Fig.~\ref{cluster_ele}) is  identical to the folding pathways at both $T=300K$ and $335K$. 

The coincidence of equilibrium and kinetic intermediates is not without precedence. A series of insightful NMR experiments have established that during the folding of apomyoglobin a kinetic intermediate is populated \cite{Jennings93Science} that has the same structure as the one characterized at equilibrium \cite{Hughson90Science}. Our study leads to the prediction that the network of states accessed kinetically are also found in folding trajectories at equilibrium. This prediction can be tested using NMR experiments as a function of pH just was done for apomyoglobin.
 
\begin{figure}
\includegraphics[width=6.in]{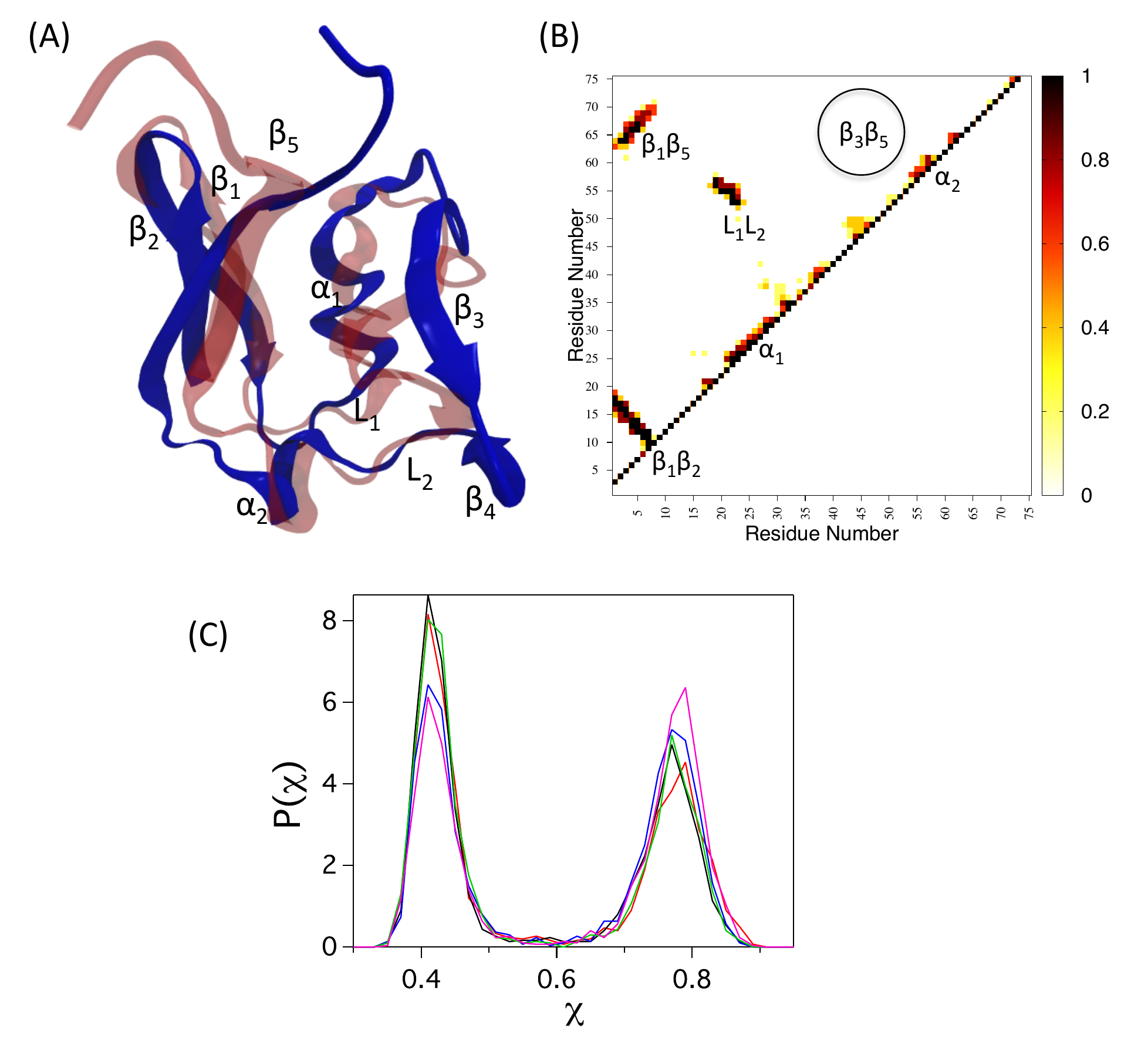}
\caption{(A) Two superimposed representative structures from the transition state ensemble at low pH. (B) The upper diagonal of the plot shows $C_{\alpha}$ contact-map of the transition state ensemble. There are no interactions between $\beta_3 \beta_5$ (shown in circle). The lower diagonal of the plot shows experimental\cite{Krantz04JMB} $\Psi$-values. (C) The distribution of the final structural overlap parameter, $\chi$, of at least 500 simulation trajectories spawned from the transition state structures.  Data is shown for five different structures. The distribution shows that roughly half of these trajectories go to the folded basin and the other half reach unfolded basin ($p_{fold} \approx 0.5$).}\label{trans}
\end{figure}

\begin{figure}
\includegraphics[width=6.5in]{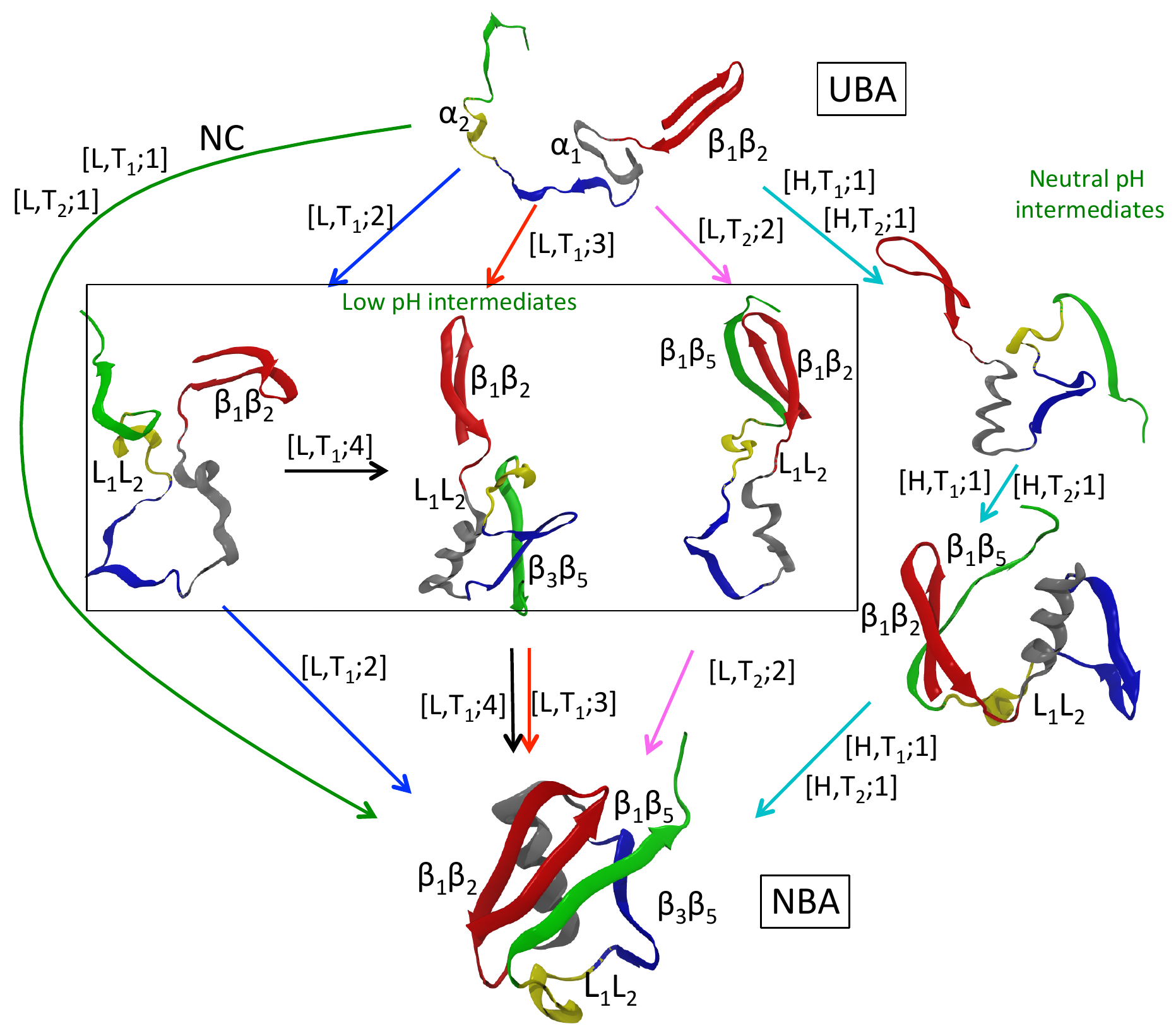}
\caption{Complex folding pH and temperature-dependent folding pathways for Ub. The first index in  [L,$T_1$; $\alpha$] denotes low pH, the second is temperature, and $\alpha$ = 1, 2, 3, 4 labels the kinetic pathways. A similar interpretation with $T_2 > T_1$ holds for [H,$T_2$; $\alpha$] with H standing for neutral pH.  In the unfolded state  secondary structures stabilized by local contacts ($\alpha_1$, $\alpha_2$, $\beta_1\beta_2$) are always present. The transition from UBA to NBA in [L,$T_1$,1] and [L,$T_2$,1] (green arrow) is described by the nucleation-collapse (NC) mechanism. Blue, red, and black arrows routes to the NBA from UBA  at $T_1$ through well-defined intermediates whose structures are displayed. The high temperature route through an intermediate is shown by lavender arrows. The folding pathway at neutral pH at both temperatures is in aqua blue. Three of the four intermediates (the left two and the one on the right) are also populated in equilibrium folding trajectories. }\label{schematic}
\end{figure}

{\bf Identification of the Transition State Ensemble (TSE) using $P_{fold}$:}
The transition state structures of Ub are identified from the folding trajectory at $T_m$ (Fig.~S2). Putative transition state structures are picked from the saddle-point region of the free energy projected onto the variables $E$ and $\chi$ using the conditions $-50.0 kcal/mole < E < -40.0 kcal/mole$ and $0.66 < \chi < 0.68$. Starting from these structures we calculated the commitment probability, $P_{fold}$\cite{Du98JCP}, of reaching the NBA.   The set of structures with  $P_{fold} \approx .4-.6$ is identified as the transition state ensemble (TSE) (Fig.~\ref{trans}C, S9). To our knowledge this is the first demonstration that TSE has been quantitatively identified for Ub without any prejudice  about the underlying reaction coordinate.  

The average TSE structure is globular  with most of the SSEs and tertiary contacts intact as in the  folded state, and they are fairly homogeneous (Fig.~\ref{trans}A) supporting the conclusions based on $\Psi$-value analysis\cite{Krantz04JMB, Krantz05JMB} and all atom simulations\cite{Piana13PNAS}. The average $\chi$, an estimate of the contact-order in the TSE, is approximately 0.67 in agreement with reported values for various proteins\cite{Paci05JMB}. The $\beta_1 \beta_2$ hairpin and $\alpha_1$ are fully structured in the TSE which is not surprising given their thermodynamic stability (Fig. 3B).  Compared to the folded structure the contacts between the $\beta$-sheets $\beta_3 \beta_5$ are absent in the TSE, although the strands $\beta_3$ and $\beta_5$ are fully structured (Fig.~\ref{trans}B). The formation of a compact TSE in which majority of the SSE and tertiary interactions are consolidated further supports that at $T \approx T_m$ Ub folds by the NC mechanism.
  
The TSE structures identified in the simulations are in reasonable agreement with the inferences drawn from the $\Phi$-value analysis\cite{Went05ProteinEng}, and are in better agreement with the $\Psi$-value analysis\cite{Sosnick04PNAS,Krantz05JMB} and $T$-jump infrared spectroscopy experiments\cite{Chung08Biochemistry}.  Based on these studies \cite{Went05ProteinEng,Sosnick04PNAS,Krantz05JMB} it is suggested that the N-terminal part of the protein, helix $\alpha_1$  and sheet $\beta_1 \beta_2$, are ordered in the TSE, which agrees with our simulations  (Fig.~\ref{trans}A).  However, the experiments disagree among each other on the TSE structure  in the C-terminal region of the protein. The $\Phi$-value analysis\cite{Went05ProteinEng} suggest that the C-terminal region of the protein is unfolded while the $\Psi$-value analysis\cite{Sosnick04PNAS,Krantz05JMB} and $T$-jump infrared spectroscopy experiments\cite{Chung08Biochemistry} infer the opposite.  According to these experiments the transition state is extensively ordered with  structure comprising the four $\beta$-strands and the $\alpha$-helix. Our simulations show that the $\beta_5$ does make contacts with $\beta_1$ in the TSE in agreement with the experiments based on $\Psi$-value analysis. In addition, the $p_{fold}$ analysis shows that the C-terminus $\alpha$-helix is at least partially structured. 

A picture of the TSE using all atom MD simulations in water and projection onto a one-dimensional reaction coordinate was proposed\cite{Piana13PNAS}. Using the dynamics in the projected coordinate   they  computed $\Phi$-values for only hydrophobic residues using certain (untested) assumptions. The trends  (not absolute values) in  experiments and simulations are similar\cite{Piana13PNAS}. On this basis they  asserted that the structures in the barrier region in the one-dimensional coordinate is the TSE.  Because of the completely different methods and  the models used (our TSE is most appropriate for acidic pH) in the two studies it is difficult to directly compare the $p_{fold}$-based determination of the TSE with the one from\cite{Piana13PNAS}. Nevertheless, in both the studies TSEs are homogeneous, compact, and native-like.

{\bf Relevance of non-native interactions:} We provide generic arguments showing that non-native interactions ought to play only a sub-dominant role in the folding of evolved small proteins that ostensibly fold in an apparent two-state manner. In order to keep the arguments simple let us assume that non-native interactions largely affect the unfolded state. There is anecdotal evidence that this is the case in a mutant of NTL9 \cite{Cho05JMB}. We write the free energy difference between the unfolded states containing non-native (NN) interactions and one described using only native (N) interactions as $\Delta \Delta G_U = \Delta H_U - T(\Delta S_U)$ where $\Delta H_U$ ($\Delta S_U$) is the enthalpy (entropy) changes between the NN and N models of the unfolded state. Typically, but not always, we expect that NN interactions ought to stabilize the unfolded state compared to the unfolded state described by the N model. However, $\Delta H_U < 0$ also implies $T(\Delta S_U)$ will be negative because certain conformations formed by favorable NN interactions are disallowed in the native interaction dominated model. Thus, the sign of $\Delta \Delta G_U$ is determined by the magnitude of $T(\Delta S_U)$, which cannot be too large to negate $\Delta H_U$. If it were the case then NN interactions alone would stabilize the folded states to a greater extent than N interactions, which is unlikely. So we will assume that $|\frac{\Delta H_U}{T(\Delta S_U)}| > 1$. 

Because hydrophobic interactions between small hydrophobic species is entropic in origin we expect that non-native interactions are most relevant when a salt bridge, not present in the folded state, can form in the unfolded state, and hence may be important in Ub folding around neutral pH.  It is clear that $|\frac{\Delta \Delta G_U}{\Delta G_{NU}}| \ll 1$ where $\Delta G_{NU}$ is the stability of the native state with respect to the unfolded state.  If this inequality is violated then the protein would not fold! Thus, it follows that NN interactions are most likely to be perturbation and not a dominant determinant of the folding thermodynamics. This conclusion is increasingly valid for proteins whose native states are highly stable. A corollary of this argument is that folding rates, and in most cases unfolding rates as well, are unlikely to change significantly (less than an order of magnitude). Our previous works on lattice models, which treated N and NN interactions on {\it equal footing}, provide illustrations of the arguments provided here.

Do the arguments given above imply that NN interactions\cite{Chen15COSB} are not relevant at all?  We discuss two examples suggesting that favorable NN interactions may affect the stability and kinetics of folding. (1) Mutations of surface exposed charged residues in Fyn SH3 domain, a small protein ($\approx 56$ residues) showed the folding rates, with respect to the WT,  increases by a factor of $\approx 3$ for the E46K mutant and by a factor of $\approx 8$ for mutant for the E46K-E11K-D16K-H21K-N30K (Fyn5) mutant (see Table I in\cite{Zarrine-Afsar12Proteins}). The unfolding rate decreases only by a factor of two for these mutants, which can be taken to mean that the effects of the dramatic mutations affect largely the unfolded states.  Even though these mutations, especially the Fyn5 construct, are drastic the effect on the folding kinetics is modest and the factor of eight increase can be accounted for by $\approx 2K_BT$ change in the barrier height, which mirrors roughly the enhanced stability of the Fyn5 mutant (see Table I in\cite{Zarrine-Afsar12Proteins}). The relatively small  changes (less than a factor of ten even in Fyn5 mutant) in the rates are in accord with the arguments given above. To explain these changes coarse-grained simulations were performed using $C_{\alpha}$ representation of the protein. We do not believe the explanation based on these simulations is adequate for two reasons. First, in these simulations included only NN interactions between charged residues using the Debye-Huckel potential.  The parameters of the NN electrostatic interactions are very different from those used for native electrostatic interactions. In other words, NN and N interactions are not treated on {\it equal footing}. Second, the inferences that non-native interactions might be present in the transition state were made based on  free energy profiles computed using the fraction of native contacts without the benefit of $P_{fold}$ analysis.  (2) In a more compelling case, it has been shown in a number of NTL9 mutants the unfolded state may be stabilized by a non-native salt bridge\cite{Cho05JMB}. The most dramatic change occurs in experiments at pH 5.5 in the K12M mutant (a rather large perturbation)  in which the free energy of the mutant is enhanced by $\approx 3K_BT$ (a 50\% increase) relative to the WT, with all other mutants exhibiting less than $K_BT$ change (see Table 1 in\cite{Cho05JMB}).  Apparently in the WT a salt bridge forms between D8K12 in the unfolded state, which is clearly abolished in the mutant, thus increasing the free energy of the unfolded state.  The result is the K12M is more stable than the WT.  These results were explained by using CG simulations in which Debye-Huckel potential was used with NN interactions only between charged residues\cite{Azia09JMB}. In this study NN and N electrostatic interactions were treated on equal footing. The findings corroborate the experimental observations. 

We draw two conclusions fro the discussion presented here. First, only in dramatically altered sequences NN  interactions are significant in affecting the stability.  Second, these changes involving charged residues can be taken into account within CG models by slightly altering the strength of hydrophobic interactions. After all large perturbations in both Fyn SH3 and NTL9 can modulate both electrostatic as well hydrophobic interactions because one expects changes in hydration in the unfolded states as a consequence of these mutations. Based on the current evidence from simulations\cite{Camacho95Proteins,Klimov01Proteins,Best13PNAS}, we conclude that generically non-native interactions are likely to be {\it only a perturbation} and {\it not} a dominant factor in the folding of small single domain proteins. This conclusion is in accord with the arguments given above and theoretical considerations\cite{Bryngelson95Proteins}.

\bigskip

{\bf Conclusions:}

In summary, using coarse-grained models and molecular dynamics simulations we have dissected the folding of Ub as a function of temperature at acidic and neutral pH. The major findings in this work are: (1) We predict quantitively the pH-dependent changes in the radius of gyration of Ub. The values of mean $R_g$ at high temperatures are in excellent agreement with  experiments.  (2) A significant prediction of our study is that the folding pathways can be dramatically altered by changing pH. The major pathway at low pH resembles the minor pathway at neutral pH.  The structures of some of the intermediates and transition states, which are only indirectly inferred from experiments,  are fully resolved.  (3) Our work also highlights the balance between the number of local and non-local contacts determines the folding mechanisms of protein folding in general\cite{Abkevich95JMB}. In the context of Ub, the secondary structural elements stabilized by local contacts form in the early stages of the folding process. Only subsequently and after considerable compaction, secondary structures, stabilized by non-local contacts, form. The formation of these non-local contacts determine the folding rates and are strongly influenced by the folding conditions. (4) Although there are dominant folding pathways under all conditions for folding in general, and Ub in particular, self-assembly can occur by alternate sub-dominant routes. Thus, the assembly mechanism of proteins should be described in probabilistic terms - a notion that appears naturally in the statistical mechanical description of folding \cite{Bryngelson95Proteins,Thirumalai95JPhysI}. In accord with this general principle, we find that the folding mechanism is complex especially at acidic pH. Under these conditions, a fraction of molecules  folds by a nucleation-collapse mechanism where as in others long lived meta-stable intermediates are populated prior to collapse and the formation of the native-state. This finding is in line with the KPM\cite{Guo95Biopolymers}, which is now firmly established for a number of proteins\cite{Peng08PNAS,Stigler11Science}. At neutral pH, Ub folds by a sequential mechanism in which local SSEs first form. Subsequently an intermediate stabilized by long range contacts ($L_1 L_2$ and $\beta_1 \beta_5$) is populated prior to the formation of the native-state. 

An important enterprise in molecular simulations is to benchmark forcefields, which should be done by comparing simulations and experiments.  Minimally such comparisons  should include specific heat profiles, dependence of the dimensions ($R_g$s)  of the protein as a function of temperature and denaturants, and time dependent changes in $R_g$ and other measurable properties probing the kinetics of self-assembly. We hasten to add that it is almost impossible to calculate accurately (nor should one attempt such computations) material-dependent properties (specific heat being one example) using simulations with {\it ad hoc} empirical force-fields including CG models or atomically detailed models.  For purposes of direct comparisons with experiments it is prudent to create transferable CG models\cite{Davtyan12JPCB,Hyeon06Structure} by benchmarking against experiments. The transferable CG force-field we have created has been remarkably successful in semi-quantitatively reproducing many experimental quantities for srcSH3 \cite{Liu11PNAS} and GFP \cite{Reddy12PNAS} as a function of denaturants. Such simulations are currently beyond the scope of atomic detailed simulations because of lack of reasonable force fields for denaturants and the sheer size of GFP.

Despite the ability to reproduce experimental measurements and make testable predictions for a large number of proteins using coarse-grained models they have obvious limitations. The absence of explicit inclusion of the solvent, which has  an impact on the fluctuations of the unfolded state, makes it difficult to quantitively reproduce the measured heat capacity curves.   Finally the knowledge of the native structure needed in these simulations can be legitimately criticized.  Despite these reservations the potential utility of coarse-grained models in protein and RNA folding is substantial\cite{Hyeon11NatComm}. Most importantly, such simulations can be carried out using standard desktop computers.
\bigskip

{\bf Acknowledgement:} 
We are indebted to George Makhadatze for pointing out an error in the simulated heat capacity curves in an earlier version. We thank William Eaton, Stefano Piana, Eugene Shaknovich, and Tobin Sosnick for valuable comments. We also acknowledge discussions with Koby Levy on non-native electrostatic interactions in NTL9. GR acknowledges startup grant from Indian Institute of Science-Bangalore. DT acknowledges a grant from the National Science Foundation through grant CHE 1361946. A portion of this research used resources of the National Energy Research Scientific Computing Center, a DOE Office of Science User Facility supported by the Office of Science of the U.S. Department of Energy under Contract No. DE-AC02-05CH11231.

\newpage

\bibliography{Proteins11}

\begin{thebibliography}{100}

\bibitem{Schuler08COSB}
Benjamin Schuler and William~A. Eaton.
\newblock {Protein folding studied by single-molecule FRET}.
\newblock {\em {Curr. Opin. Struct. Biol.}}, {18}({1}):{16--26}, {2008}.

\bibitem{vzoldak13COSB}
Gabriel {\v{Z}}old{\'a}k and Matthias Rief.
\newblock Force as a single molecule probe of multidimensional protein energy
  landscapes.
\newblock {\em {Curr. Opin. Struct. Biol.}}, 23(1):48--57, 2013.

\bibitem{Wolynes95Science}
P~G Wolynes, J~N Onuchic, and D~Thirumalai.
\newblock {Navigating the Folding Routes}.
\newblock {\em {Science}}, {267}({5204}):{1619--1620}, {MAR 17} {1995}.

\bibitem{Bryngelson95Proteins}
J~D Bryngelson, J~N Onuchic, N~D Socci, and P~G Wolynes.
\newblock {Funnels, pathways, and the energy landscape of protein folding: A
  synthesis}.
\newblock {\em {Proteins}}, {21}({3}):{167--195}, {MAR} {1995}.

\bibitem{Dill97NSB}
KA~Dill and HS~Chan.
\newblock {From Levinthal to pathways to funnels}.
\newblock {\em {Nat. Struct. Biol.}}, {4}({1}):{10--19}, {1997}.

\bibitem{Thirumalai99COSB}
D~Thirumalai and DK~Klimov.
\newblock {Deciphering the timescales and mechanisms of protein folding using
  minimal off-lattice models}.
\newblock {\em {Curr. Opin. Struct. Biol.}}, {9}({2}):{197--207}, {1999}.

\bibitem{Shakhnovich06ChemRev}
E~Shakhnovich.
\newblock {Protein folding thermodynamics and dynamics: Where physics,
  chemistry, and biology meet}.
\newblock {\em {Chem. Rev.}}, {106}:{1559--1588}, {2006}.

\bibitem{Thirumalai10ARB}
D.~Thirumalai, E.~P. O'Brien, G.~Morrison, and C.~Hyeon.
\newblock {Theoretical Perspectives on Protein Folding}.
\newblock {\em {Ann. Rev. Biophys.}}, {39}:{159--183}, {2010}.

\bibitem{Dill12Science}
Ken~A. Dill and Justin~L. MacCallum.
\newblock {The Protein-Folding Problem, 50 Years On}.
\newblock {\em {Science}}, {338}({6110}):{1042--1046}, {2012}.

\bibitem{Hyeon11NatComm}
Changbong Hyeon and D.~Thirumalai.
\newblock {Capturing the essence of folding and functions of biomolecules using
  coarse-grained models}.
\newblock {\em {Nat. Commun.}}, {2}:{487}, {2011}.

\bibitem{Whitford12RepProg}
Paul~C. Whitford, Karissa~Y. Sanbonmatsu, and Jose~N. Onuchic.
\newblock {Biomolecular dynamics: order-disorder transitions and energy
  landscapes}.
\newblock {\em {Rep. Prog. Phys.}}, {75}({7}), {JUL} {2012}.

\bibitem{Tozzini10QRB}
Valentina Tozzini.
\newblock {Minimalist models for proteins: a comparative analysis}.
\newblock {\em {Q. Rev. Biophys.}}, {43}({3}):{333--371}, {2010}.

\bibitem{Vicatos14Proteins}
Spyridon Vicatos, Anna Rychkova, Shayantani Mukherjee, and Arieh Warshel.
\newblock {An effective Coarse-grained model for biological simulations: Recent
  refinements and validations}.
\newblock {\em {Proteins}}, {82}({7}):{1168--1185}, {2014}.

\bibitem{Best13PNAS}
Robert~B. Best, Gerhard Hummer, and William~A. Eaton.
\newblock {Native contacts determine protein folding mechanisms in atomistic
  simulations}.
\newblock {\em {Proc. Natl. Acad. Sci. U. S. A.}}, {110}({44}):{17874--17879},
  {OCT 29} {2013}.

\bibitem{Shaw10Science}
D.~E. Shaw, P.~Maragakis, K.~Lindorff-Larsen, S.~Piana, R.~O. Dror, M.~P.
  Eastwood, J.~A. Bank, J.~M. Jumper, J.~K. Salmon, Y.~Shan, and W.~Wriggers.
\newblock {Atomic-Level Characterization of the Structural Dynamics of
  Proteins}.
\newblock {\em {Science}}, {330}:{341--346}, {2010}.

\bibitem{Lindorff11Science}
Kresten Lindorff-Larsen, Stefano Piana, Ron~O Dror, and David~E Shaw.
\newblock How fast-folding proteins fold.
\newblock {\em Science}, 334(6055):517--520, 2011.

\bibitem{Thirumalai13COSB}
D~Thirumalai, Zhenxing Liu, Edward~P O'Brien, and Govardhan Reddy.
\newblock {Protein folding: From theory to practice}.
\newblock {\em {Curr. Opin. Struct. Biol.}}, 23(1):22--29, 2013.

\bibitem{Hicke01NatRevMolCell}
L~Hicke.
\newblock Protein regulation by monoubiquitin.
\newblock {\em {Nat. Rev. Mol. Cell. Biol.}}, 2:195--201, 2001.

\bibitem{Finley09ARBiochem}
D~Finley.
\newblock Recognition and processing of ubiquitin-protein conjugates by the
  proteasome.
\newblock {\em Ann. Rev. Biochem.}, 78:477--513, 2009.

\bibitem{Lee14JMB}
S.~Y. Lee, L~Pullen, Daniel~J. Virgi, I, C.~A. Castaneda, D.~Abeykoon, D.~N.~A.
  Bolon, and D~Fushman.
\newblock {Alanine Scan of Core Positions in Ubiquitin Reveals Links between
  Dynamics, Stability, and Function}.
\newblock {\em {J. Mol. Biol.}}, 426(7):1377--1389, 2014.

\bibitem{Thirumalai95JPhysI}
D~Thirumalai.
\newblock {From Minimal Models to Real Proteins: Time Scales for Protein
  Folding Kinetics}.
\newblock {\em {J. Phys. I}}, {5}({11}):{1457--1467}, {1995}.

\bibitem{Khorasanizadeh96NSB}
S~Khorasanizadeh, ID~Peters, and H~Roder.
\newblock {Evidence for a three-state model of protein folding from kinetic
  analysis of ubiquitin variants with altered core residues}.
\newblock {\em {Nat Struct Biol}}, {3}({2}):{193--205}, {1996}.

\bibitem{Chung08Biochemistry}
Hoi~Sung Chung, Ali Shandiz, Tobin~R. Sosnick, and Andrei Tokmakoff.
\newblock {Probing the Folding Transition State of Ubiquitin Mutants by
  Temperature-Jump-Induced Downhill Unfolding}.
\newblock {\em {Biochemistry}}, {47}({52}):{13870--13877}, {2008}.

\bibitem{Hyeon06Structure}
C.B. Hyeon, R.~I. Dima, and D.~Thirumalai.
\newblock Pathways and kinetic barriers in mechanical unfolding and refolding
  of {RNA} and proteins.
\newblock {\em Structure}, 14:1633--1645, {2006}.

\bibitem{Reddy12PNAS}
Govardhan Reddy, Zhenxing Liu, and D.~Thirumalai.
\newblock {Denaturant-dependent folding of GFP}.
\newblock {\em {Proc. Natl. Acad. Sci. USA}}, {109}:{17832--17838}, 2012.

\bibitem{Chen10PNAS}
Jie Chen, Seth~A. Darst, and D.~Thirumalai.
\newblock {Promoter melting triggered by bacterial RNA polymerase occurs in
  three steps}.
\newblock {\em {Proc. Natl. Acad. Sci. U. S. A.}}, {107}({28}):{12523--12528},
  {2010}.

\bibitem{Theisen12JPCB}
K.~E. Theisen, A~Zhmurov, M.~E. Newberry, V~Barsegov, and R.~I. Dima.
\newblock {Multiscale Modeling of the Nanomechanics of Microtubule
  Protofilaments}.
\newblock {\em {J. Phys. Chem. B}}, 116:8545--8555, 2012.

\bibitem{Kononova14JACS}
O~Kononova, Y~Kholodov, K.~E. Theisen, K.~A. Marx, R.~I. Dima, F.~I.
  Ataullakhanov, E.~L. Grishchuk, and V~Barsegov.
\newblock {Tubulin Bond Energies and Microtubule Biomechanics Determined from
  Nanoindentation in Silico}.
\newblock {\em {J. Am. Chem. Soc.}}, 136(49):17036--17045, 2014.

\bibitem{Kononova13BJ}
O~Kononova, J~Snijder, M~Brasch, J~Cornelissen, R.~I. Dima, K.~A. Marx,
  G.~J.~L. Wuite, W.~H. Roos, and V~Barsegov.
\newblock {Structural Transitions and Energy Landscape for Cowpea Chlorotic
  Mottle Virus Capsid Mechanics from Nanomanipulation in Vitro and in Silico}.
\newblock {\em {Biophys. J. }}, 105(8):1893--1903, 2013.

\bibitem{Camacho95Proteins}
C.J. Camacho and D.~Thirumalai.
\newblock {Modeling the role of disulfide bonds in protein folding: Entropic
  barriers and pathways}.
\newblock {\em Proteins: Structure, Function, and Bioinformatics},
  22(1):27--40, 1995.

\bibitem{Klimov01Proteins}
D.~Klimov and D.~Thirumalai.
\newblock Multiple protein folding nuclei and the transition state ensemble in
  two state proteins.
\newblock {\em Proteins Struct. Funct. Gen.}, 43:465--475, 2001.

\bibitem{Fernandez02Physica}
A~Fernandez, A~Colubri, and RS~Berry.
\newblock {Three-body correlations in protein folding: the origin of
  cooperativity}.
\newblock {\em {Physica A}}, {307}({1-2}):{235--259}, {2002}.

\bibitem{Marianayagam04BiophysChem}
NJ~Marianayagam and SE~Jackson.
\newblock {The folding pathway of ubiquitin from all-atom molecular dynamics
  simulations}.
\newblock {\em {Biophys Chem}}, {111}({2}):{159--171}, {2004}.

\bibitem{Alonso95JMB}
DOV Alonso and V~Daggett.
\newblock {Molecular dynamics simulations of protein unfolding and limited
  refolding - characterization of partially unfolded states of ubiquitin in
  60-percent methnol and in water}.
\newblock {\em {J. Mol. Biol.}}, {247}({3}):{501--520}, {1995}.

\bibitem{Alonso98ProteinScience}
DOV Alonso and V~Daggett.
\newblock {Molecular dynamics simulations of hydrophobic collapse of
  ubiquitin}.
\newblock {\em {Protein Sci}}, {7}({4}):{860--874}, {1998}.

\bibitem{Irback06Proteins}
Anders Irback and Simon Mitternacht.
\newblock {Thermal versus mechanical unfolding of ubiquitin}.
\newblock {\em {Proteins}}, {65}({3}):{759--766}, {2006}.

\bibitem{Sorensen02Proteins}
JM~Sorensen and T~Head-Gordon.
\newblock {Toward minimalist models of larger proteins: A ubiquitin-like
  protein}.
\newblock {\em {Proteins}}, {46}({4}):{368--379}, {2002}.

\bibitem{Dastidar05PRE}
SG~Dastidar and C~Mukhopadhyay.
\newblock {Unfolding dynamics of the protein ubiquitin: Insight from
  simulation}.
\newblock {\em {Phys. Rev. E}}, {72}({5}):{051928}, {2005}.

\bibitem{Kony07ProteinScience}
David~B. Kony, Philippe~H. Hunenberger, and Wilfred~F. van Gunsteren.
\newblock {Molecular dynamics simulations of the native and partially folded
  states of ubiquitin: Influence of methanol cosolvent, pH, and temperature on
  the protein structure and dynamics}.
\newblock {\em {Protein Sci}}, {16}({6}):{1101--1118}, {2007}.

\bibitem{Zhang05Proteins}
J~Zhang, M~Qin, and W~Wang.
\newblock {Multiple folding mechanisms of protein ubiquitin}.
\newblock {\em {Proteins}}, {59}({3}):{565--579}, {2005}.

\bibitem{Piana13PNAS}
Stefano Piana, Kresten Lindorff-Larsen, and David~E. Shaw.
\newblock {Atomic-Level Description of Ubiquitin Folding}.
\newblock {\em {Proc. Natl. Acad. Sci. U. S. A.}}, {110}({15}):{5915--5920},
  {2013}.

\bibitem{Mandal14PCCP}
Manoj Mandal and Chaitali Mukhopadhyay.
\newblock {Microsecond molecular dynamics simulation of guanidinium chloride
  induced unfolding of ubiquitin}.
\newblock {\em {Phys. Chem. Chem. Phys.}}, {16}({39}):{21706--21716}, {2014}.

\bibitem{Sorenson02Proteins}
{Sorenson, J. M. and Head-Gordon, T}.
\newblock {Toward Minimalist Models of Larger Proteins: a
  Ubiquitin-likeProtein}.
\newblock {\em Proteins}, 46(4):368--379, 2002.

\bibitem{Wintrode94Proteins}
PL~Wintrode, GI~Makhatadze, and PL~Privalov.
\newblock {Thermodynamics of ubiquitin unfolding}.
\newblock {\em {Proteins}}, {18}({3}):{246--253}, {1994}.

\bibitem{Betancourt99ProtSci}
M.R. Betancourt and D.~Thirumalai.
\newblock Pair potentials for protein folding: {C}hoice of reference states and
  sensitivity of predicted native states to variations in the interaction
  schemes.
\newblock {\em Prot. Sci.}, 8:361--369, 1999.

\bibitem{Kumar87JMB}
S~Vjaykumar, CE~Bugg, and WJ~Cook.
\newblock {Structure of ubiquitin refined at 1.8 A resolution}.
\newblock {\em {J. Mol. Biol.}}, {194}({3}):{531--544}, {1987}.

\bibitem{Fogolari03Biophys}
F~Fogolari, A~Brigo, and H~Molinari.
\newblock {Protocol for MM/PBSA molecular dynamics simulations of proteins}.
\newblock {\em {Biophys. J.}}, {85}({1}):{159--166}, {2003}.

\bibitem{Kurnik12PNAS}
Martin Kurnik, Linda Hedberg, Jens Danielsson, and Mikael Oliveberg.
\newblock {Folding without charges}.
\newblock {\em {Proc. Natl. Acad. Sci. U. S. A.}}, {109}({15}):{5705--5710},
  {2012}.

\bibitem{Cardenas03Proteins}
A.E. Cardenas and R.~Elber.
\newblock Kinetics of cytochrome {C} folding: {Atomically} detailed
  simulations.
\newblock {\em Proteins Struct. Funct. Gen.}, 51:245--257, 2003.

\bibitem{Li00NSB}
L~Li, LA~Mirny, and EI~Shakhnovich.
\newblock {Kinetics, thermodynamics and evolution of non-native interactions in
  a protein folding nucleus}.
\newblock {\em {Nat. Struct. Biol. }}, 7(4):336--342, 2000.

\bibitem{Liu11PNAS}
Zhenxing Liu, Govardhan Reddy, Edward~P O'Brien, and D~Thirumalai.
\newblock Collapse kinetics and chevron plots from simulations of
  denaturant-dependent folding of globular proteins.
\newblock {\em {Proc. Natl. Acad. Sci. USA}}, 108(19):7787--7792, 2011.

\bibitem{Veitshans97FoldDes}
T.~Veitshans, D.~Klimov, and D.~Thirumalai.
\newblock Protein folding kinetics: Timescales, pathways and energy landscapes
  in terms of sequence-dependent properties.
\newblock {\em Fold Des}, 2(1):1--22, 1997.

\bibitem{Ermak78JChemPhys}
D.~L. Ermak and J.~A. Mccammon.
\newblock Brownian dynamics with hydrodynamic interactions.
\newblock {\em J Chem Phys}, 69(4):1352--1360, 1978.

\bibitem{Guo96JMB}
Z.~Guo and D.~Thirumalai.
\newblock Kinetics and thermodynamics of folding of a de novo designed four
  helix bundle.
\newblock {\em J. Mol. Biol.}, 263:323--343, 1996.

\bibitem{Zhou12JCTC}
Ting Zhou and Amedeo Caflisch.
\newblock Distribution of reciprocal of interatomic distances: A fast
  structural metric.
\newblock {\em J. Chem. Theory Comput.}, 8(8):2930--2937, 2012.

\bibitem{Seeber07Bioinfo}
Michele Seeber, Marco Cecchini, Francesco Rao, Giovanni Settanni, and Amedeo
  Caflisch.
\newblock Wordom: a program for efficient analysis of molecular dynamics
  simulations.
\newblock {\em Bioinformatics}, 23(19):2625--2627, 2007.

\bibitem{Spath80Cluster}
Helmuth Sp{\"a}th.
\newblock {\em Cluster analysis algorithms for data reduction and
  classification of objects}.
\newblock Horwood, 1980.

\bibitem{Molero99Biochemistry}
B~Ibarra-Molero, VV~Loladze, GI~Makhatadze, and JM~Sanchez-Ruiz.
\newblock {Thermal versus guanidine-induced unfolding of ubiquitin. An analysis
  in terms of the contributions from charge-charge interactions to protein
  stability}.
\newblock {\em {Biochemistry}}, {38}({25}):{8138--8149}, {1999}.

\bibitem{Zhou99ProtSci}
Yaoqi Zhou, Carol~K Hall, and Martin Karplus.
\newblock The calorimetric criterion for a two-state process revisited.
\newblock {\em Prot. Sci.}, 8(5):1064--1074, 1999.

\bibitem{Huang12JACS}
Jie-rong Huang, Frank Gabel, Malene~Ringkjobing Jensen, Stephan Grzesiek, and
  Martin Blackledge.
\newblock {Sequence-Specific Mapping of the Interaction between Urea and
  Unfolded Ubiquitin from Ensemble Analysis of NMR and Small Angle Scattering
  Data}.
\newblock {\em {J. Am. Chem. Soc.}}, {134}({9}):{4429--4436}, {2012}.

\bibitem{Gabel09JACS}
Frank Gabel, Malene~Ringkjobing Jensen, Giuseppe Zaccai, and Martin Blackledge.
\newblock {Quantitative Modelfree Analysis of Urea Binding to Unfolded
  Ubiquitin Using a Combination of Small Angle X-ray and Neutron Scattering}.
\newblock {\em {J. Am. Chem. Soc.}}, {131}({25}):{8769--8771}, {2009}.

\bibitem{Jacob04JMB}
J~Jacob, B~Krantz, RS~Dothager, P~Thiyagarajan, and TR~Sosnick.
\newblock {Early collapse is not an obligate step in protein folding}.
\newblock {\em {J. Mol. Biol.}}, {338}({2}):{369--382}, {2004}.

\bibitem{Candotti13PNAS}
Michela Candotti, Santiago Esteban-Martin, Xavier Salvatella, and Modesto
  Orozco.
\newblock {Toward an atomistic description of the urea-denatured state of
  proteins}.
\newblock {\em {Proc. Natl. Acad. Sci. U. S. A.}}, {110}({15}):{5933--5938},
  {2013}.

\bibitem{Piana14COSB}
Stefano Piana, John~L. Klepeis, and David~E. Shaw.
\newblock {Assessing the accuracy of physical models used in protein-folding
  simulations: quantitative evidence from long molecular dynamics simulations}.
\newblock {\em {Curr. Opin. Struct. Biol.}}, {24}:{98--105}, {2014}.

\bibitem{Skinner14PNAS}
John~J Skinner, Wookyung Yu, Elizabeth~K Gichana, Michael~C Baxa, James~R
  Hinshaw, Karl~F Freed, and Tobin~R Sosnick.
\newblock Benchmarking all-atom simulations using hydrogen exchange.
\newblock {\em {Proc. Natl. Acad. Sci. U. S. A.}}, 111(45):15975--15980, 2014.

\bibitem{Best14JCTC}
Robert~B. Best, Wenwei Zheng, and Jeetain Mittal.
\newblock {Balanced Protein-Water Interactions Improve Properties of Disordered
  Proteins and Non-Specific Protein Association}.
\newblock {\em {J. Chem. Theory Comput.}}, {10}({11}):{5113--5124}, {NOV}
  {2014}.

\bibitem{Piana15JPCB}
Stefano Piana, Alexander~G. Donchev, Paul Robustelli, and David~E. Shaw.
\newblock Water dispersion interactions strongly influence simulated structural
  properties of disordered protein states.
\newblock {\em The Journal of Physical Chemistry B}, 119(16):5113--5123, 2015.

\bibitem{Camacho93PRL}
C.~J. Camacho and D.~Thirumalai.
\newblock Minimum energy compact structures of random sequences of
  heteropolymers.
\newblock {\em Phys. Rev. Lett.}, 71:2505--2508, 1993.

\bibitem{GarciaManyes09PNAS}
Sergi Garcia-Manyes, Lorna Dougan, Carmen~L. Badilla, Jasna Brujic, and
  Julio~M. Fernandez.
\newblock {Direct observation of an ensemble of stable collapsed states in the
  mechanical folding of ubiquitin}.
\newblock {\em {Proc. Natl. Acad. Sci.}}, 106(26):10534--10539, 2009.

\bibitem{Krantz04JMB}
BA~Krantz, RS~Dothager, and TR~Sosnick.
\newblock {Discerning the structure and energy of multiple transition states in
  protein folding using psi-analysis}.
\newblock {\em {J. Mol. Biol.}}, {337}({2}):{463--475}, {2004}.

\bibitem{Krantz05JMB}
BA~Krantz, RS~Dothager, and TR~Sosnick.
\newblock {Discerning the structure and energy of multiple transition states in
  protein folding using psi-analysis (vol 337, pg 463, 2004)}.
\newblock {\em {J. Mol. Biol.}}, {347}({5}):{1103}, {2005}.

\bibitem{Guo95Biopolymers}
Z.~Y. Guo and D.~Thirumalai.
\newblock Kinetics of protein-folding - nucleation mechanism, time scales, and
  pathways.
\newblock {\em Biopolymers}, 36(1):83--102, 1995.

\bibitem{Peng08PNAS}
Qing Peng and Hongbin Li.
\newblock {Atomic force microscopy reveals parallel mechanical unfolding
  pathways of T4 lysozyme: Evidence for a kinetic partitioning mechanism}.
\newblock {\em {Proc. Natl. Acad Sci.}}, 105(6):1885--1890, 2008.

\bibitem{Stigler11Science}
Johannes Stigler, Fabian Ziegler, Anja Gieseke, J.~Christof~M. Gebhardt, and
  Matthias Rief.
\newblock {The Complex Folding Network of Single Calmodulin Molecules}.
\newblock {\em {Science}}, {334}({6055}):{512--516}, {2011}.

\bibitem{Karplus79Biopolymers}
M~Karplus and DL~Weaver.
\newblock {Diffusion-Collision model for protein folding}.
\newblock {\em {Biopolymers}}, {18}({6}):{1421--1437}, {1979}.

\bibitem{Belisle12NSB}
Alexis Vallee-Belisle and Stephen~W. Michnick.
\newblock {Visualizing transient protein-folding intermediates by
  tryptophan-scanning mutagenesis}.
\newblock {\em {Nat Struct Mol Biol}}, {19}({7}):{731+}, {2012}.

\bibitem{Rea08Biochemistry}
Anita~M. Rea, Emma~R. Simpson, Jill~K. Meldrum, Huw E.~L. Williams, and Mark~S.
  Searle.
\newblock {Aromatic Residues Engineered into the beta-Turn Nucleation Site of
  Ubiquitin Lead to a Complex Folding Landscape, Non-Native Side-Chain
  Interactions, and Kinetic Traps}.
\newblock {\em {Biochemistry}}, {47}({48}):{12910--12922}, {2008}.

\bibitem{Schanda07PNAS}
Paul Schanda, Vincent Forge, and Bernhard Brutscher.
\newblock {Protein folding and unfolding studied at atomic resolution by fast
  two-dimensional NMR spectroscopy}.
\newblock {\em {Proc. Natl. Acad. Sci. U. S. A.}}, {104}({27}):{11257--11262},
  {2007}.

\bibitem{Belisle07JMB}
Alexis Vallee-Belisle and Stephen~W. Michnick.
\newblock {Multiple tryptophan probes reveal that ubiquitin folds via a late
  misfolded intermediate}.
\newblock {\em {J. Mol. Biol.}}, {374}({3}):{791--805}, {2007}.

\bibitem{Crespo06JMB}
Maria~D. Crespo, Emma~R. Simpson, and Mark~S. Searle.
\newblock {Population of on-pathway intermediates in the folding of ubiquitin}.
\newblock {\em {J. Mol. Biol.}}, {360}({5}):{1053--1066}, {2006}.

\bibitem{Larios04JMB}
E~Larios, JS~Li, K~Schulten, H~Kihara, and M~Gruebele.
\newblock {Multiple probes reveal a native-like intermediate during
  low-temperature refolding of ubiquitin}.
\newblock {\em {J. Mol. Biol.}}, {340}({1}):{115--125}, {2004}.

\bibitem{Went04FEBS}
HM~Went, CG~Benitez-Cardoza, and SE~Jackson.
\newblock {Is an intermediate state populated on the folding pathway of
  ubiquitin?}
\newblock {\em {FEBS Lett.}}, {567}({2-3}):{333--338}, {2004}.

\bibitem{Kitahara03PNAS}
R~Kitahara and K~Akasaka.
\newblock {Close identity of a pressure-stabilized intermediate with a kinetic
  intermediate in protein folding}.
\newblock {\em {Proc. Natl. Acad. Sci. U. S. A.}}, {100}({6}):{3167--3172},
  {2003}.

\bibitem{Qin02JPCB}
Z~Qin, J~Ervin, E~Larios, M~Gruebele, and H~Kihara.
\newblock {Formation of a compact structured ensemble without fluorescence
  signature early during ubiquitin folding}.
\newblock {\em {J. Phys. Chem. B}}, {106}({50}):{13040--13046}, {2002}.

\bibitem{Cordier02JMB}
F~Cordier and S~Grzesiek.
\newblock {Temperature-dependence properties as studied by of protein hydrogen
  bond high-resolution NMR}.
\newblock {\em {J. Mol. Biol.}}, {317}({5}):{739--752}, {2002}.

\bibitem{Khorasanizadeh93Biochemistry}
S~Khorasanizadeh, ID~Peters, TR~Butt, and H~Roder.
\newblock {Folding and stability of a tryptophan-containing mutant of
  ubiquitin}.
\newblock {\em {Biochemistry}}, {32}({27}):{7054--7063}, {1993}.

\bibitem{Briggs92PNAS}
MS~Briggs and H~Roder.
\newblock {Early hydrogen-bonding events in the folding reaction of ubiquitin}.
\newblock {\em {Proc. Natl. Acad. Sci. U. S. A.}}, {89}({6}):{2017--2021},
  {1992}.

\bibitem{Chung05PNAS}
HS~Chung, M~Khalil, AW~Smith, Z~Ganim, and A~Tokmakoff.
\newblock {Conformational changes during the nanosecond-to-millisecond
  unfolding of ubiquitin}.
\newblock {\em {Proc. Natl. Acad. Sci. U. S. A.}}, {102}({3}):{612--617},
  {2005}.

\bibitem{Zheng10JMB}
Zhongzhou Zheng and Tobin~R. Sosnick.
\newblock {Protein Vivisection Reveals Elusive Intermediates in Folding}.
\newblock {\em {J. Mol. Biol.}}, {397}({3}):{777--788}, {2010}.

\bibitem{Jennings93Science}
Patricia~A Jennings and Peter~E Wright.
\newblock Formation of a molten globule intermediate early in the kinetic
  folding pathway of apomyoglobin.
\newblock {\em Science}, 262(5135):892--896, 1993.

\bibitem{Hughson90Science}
Frederick~M Hughson, Peter~E Wright, and Robert~L Baldwin.
\newblock Structural characterization of a partly folded apomyoglobin
  intermediate.
\newblock {\em Science}, 249(4976):1544--1548, 1990.

\bibitem{Du98JCP}
R~Du, VS~Pande, AY~Grosberg, T~Tanaka, and ES~Shakhnovich.
\newblock {On the transition coordinate for protein folding}.
\newblock {\em {J. Chem. Phys.}}, {108}({1}):{334--350}, {1998}.

\bibitem{Paci05JMB}
E~Paci, K~Lindorff-Larsen, CM~Dobson, M~Karplus, and M~Vendruscolo.
\newblock {Transition state contact orders correlate with protein folding
  rates}.
\newblock {\em {J. Mol. Biol.}}, {352}({3}):{495--500}, {2005}.

\bibitem{Went05ProteinEng}
HM~Went and SE~Jackson.
\newblock {Ubiquitin folds through a highly polarized transition state}.
\newblock {\em {Protein Eng}}, {18}({5}):{229--237}, {2005}.

\bibitem{Sosnick04PNAS}
TR~Sosnick, RS~Dothager, and BA~Krantz.
\newblock {Differences in the folding transition state of ubiquitin indicated
  by phi and psi analyses}.
\newblock {\em {Proc. Natl. Acad. Sci. U. S. A.}}, {101}({50}):{17377--17382},
  {2004}.

\bibitem{Cho05JMB}
JH~Cho and DP~Raleigh.
\newblock {Mutational analysis demonstrates that specific electrostatic
  interactions can play a key role in the denatured state ensemble of
  proteins}.
\newblock {\em {J. Mol. Biol.}}, {353}({1}):{174--185}, {2005}.

\bibitem{Chen15COSB}
Tao Chen, Jianhui Song, and Hue~Sun Chan.
\newblock {Theoretical perspectives on nonnative interactions and intrinsic
  disorder in protein folding and binding}.
\newblock {\em {Curr. Opin. Struct. Biol.}}, {30}:{32--42}, {2015}.

\bibitem{Zarrine-Afsar12Proteins}
Arash Zarrine-Afsar, Zhuqing Zhang, Katrina~L. Schweiker, George~I. Makhatadze,
  Alan~R. Davidson, and Hue~Sun Chan.
\newblock {Kinetic consequences of native state optimization of surface-exposed
  electrostatic interactions in the Fyn SH3 domain}.
\newblock {\em {Proteins}}, {80}({3}):{858--870}, {2012}.

\bibitem{Azia09JMB}
Ariel Azia and Yaakov Levy.
\newblock {Nonnative Electrostatic Interactions Can Modulate Protein Folding:
  Molecular Dynamics with a Grain of Salt}.
\newblock {\em {J. Mol. Biol.}}, {393}({2}):{527--542}, {2009}.

\bibitem{Abkevich95JMB}
V~I Abkevich, A~M Gutin, and E~I Shakhnovich.
\newblock {Impact of Local and Non-local Interactions on Thermodynamics and
  Kinetics of Protein Folding}.
\newblock {\em {J. Mol. Biol.}}, {252}({4}):{460--471}, {1995}.

\bibitem{Davtyan12JPCB}
Aram Davtyan, Nicholas~P. Schafer, Weihua Zheng, Cecilia Clementi, Peter~G.
  Wolynes, and Garegin~A. Papoian.
\newblock {AWSEM-MD: Protein Structure Prediction Using Coarse-Grained Physical
  Potentials and Bioinformatically Based Local Structure Biasing}.
\newblock {\em {J. Phys. Chem. B}}, 116:8494--8503, 2012.

\bibitem{vmd}
W.~Humphrey, A.~Dalke, and K.~Schulten.
\newblock {VMD} - {V}isual {M}olecular {D}ynamics.
\newblock {\em {J. Molec. Graphics}}, 14:33--38, 1996.

\end{thebibliography}


\begin{thebibliography}{1}

\bibitem{Veitshans97FoldDes}
T.~Veitshans, D.~Klimov, and D.~Thirumalai.
\newblock Protein folding kinetics: Timescales, pathways and energy landscapes
  in terms of sequence-dependent properties.
\newblock {\em Fold Des}, 2(1):1--22, 1997.

\bibitem{Swope82JCP}
W.C. Swope, H.C. Andersen, P.H. Berens, and K.R. Wilson.
\newblock A computer simulation method for the calculation of equilibrium
  constants for the formation of physical clusters of molecules: application to
  small clusters.
\newblock {\em J. Chem. Phys.}, 76:637--649, 1982.

\bibitem{Ermak78JChemPhys}
D.~L. Ermak and J.~A. Mccammon.
\newblock Brownian dynamics with hydrodynamic interactions.
\newblock {\em J Chem Phys}, 69(4):1352--1360, 1978.

\bibitem{Camacho93PNAS}
C.~J. Camacho and D.~Thirumalai.
\newblock Kinetics and thermodynamics of folding in model proteins.
\newblock {\em Proc. Natl Acad Sci USA}, 90(13):6369--6372, 1993.

\bibitem{Wintrode94Proteins}
PL~Wintrode, GI~Makhatadze, and PL~Privalov.
\newblock {Thermodynamics of ubiquitin unfolding}.
\newblock {\em {Proteins}}, {18}({3}):{246--253}, {1994}.

\bibitem{Molero99Biochemistry}
B~Ibarra-Molero, VV~Loladze, GI~Makhatadze, and JM~Sanchez-Ruiz.
\newblock {Thermal versus guanidine-induced unfolding of ubiquitin. An analysis
  in terms of the contributions from charge-charge interactions to protein
  stability}.
\newblock {\em {Biochemistry}}, {38}({25}):{8138--8149}, {1999}.

\end{thebibliography}
\bibliographystyle{unsrt}

\newpage			

\end{document}


\setcounter{figure}{0}
\setcounter{table}{0}
\setcounter{equation}{0}
\renewcommand{\thefigure}{S\arabic{figure}}
\renewcommand{\thetable}{S\arabic{table}}
\renewcommand{\theequation}{S\arabic{equation}}
			
\title{Supporting Information \\ Dissecting Ubiquitin Folding Using the Self-Organized Polymer Model}
\author{Govardhan Reddy$^{* \dagger}$ and D. Thirumalai$^{\ddagger}$} 
\affiliation{$\dagger$Solid State and Structural Chemistry Unit, Indian Institute of Science, Bangalore, Karnataka, India 560012 \\$^\ddagger$Biophysics program, Institute for Physical Sciences and Technology, and Department of Chemistry and Biochemistry, University of Maryland, College Park, MD 20742}\date{\today}

\maketitle
\newpage

\textbf{SI Text}

\bigskip



\bigskip

{\bf Simulations:} In order to sample the conformations of the polypeptide chain efficiently we used low friction Langevin dynamics simulations\cite{Veitshans97FoldDes} to calculate the thermodynamic properties.   The equation of motion used in the Langevin dynamics simulations is $m \ddot{\vec{r_i}} = -\zeta \dot{\vec{r_i}} + \vec{F_c} + \vec{\Gamma}$, where $m$ is the mass of the protein beads, $\zeta$ is the friction coefficient, $\vec{r_i}$ is the position of the bead $i$, $\vec{F_c} = \frac{\partial E_{TOT}}{\partial {\vec{r_i}}}$, $\vec{\Gamma}$ is the random force with a white noise spectrum. In the discretized form the autocorrelation function of the random force is $\left < \Gamma (t) \ \Gamma (t+nh) \right > = \frac{2 \zeta k_B T}{h} \delta_{0,n}$, where $n = 0, 1, ...$ and $\delta_{0,n}$ is the Kronecker delta function. The Langevin equation is integrated using the velocity Verlet algorithm\cite{Veitshans97FoldDes, Swope82JCP}. We used $\zeta = 0.05 m/\tau_L$ and $h = 0.005\tau_L$, where $\tau_L$ is the unit of time to compute the thermodynamic properties.

To provide a realistic description of the folding kinetics we performed Brownian dynamics simulations and the equations of motion are integrated using the Ermak-McCammon algorithm\cite{Ermak78JChemPhys}, $\vec{r_i}(t+h) = \vec{r_i}(t) + \frac{h}{\zeta} \vec{F_c} + \vec{\Gamma}$. Here $\vec{\Gamma}$ is a random force with a Gaussian distribution with mean zero and variance $\left < \Gamma(h)^2 \right > = \frac{2k_B T h}{\zeta}$. The friction coefficient $\zeta = 50 m/\tau_H$ approximately corresponds to the value in water and  $h = 0.005\tau_H$. In the simulations, the characteristic unit of length $a= 1$\AA, energy $\epsilon= 1kcal/mole$, and mass $m=1.8\times10^{-22}g$ (typical mass of the bead). The unit of time in Langevin dynamics simulations is $\tau_L(= \sqrt{m a^2/\epsilon}) =0.51ps$. In Brownian dynamics, simulation time is mapped into real time, $\tau_H$ using  $\tau_H \approx \frac{\zeta_H a^2}{k_B T} = \frac{(\zeta_H \tau_L/m) \epsilon}{k_B T} \tau_L \approx 43 ps$.

\newpage

\bibliography{Proteins11}
\bibliographystyle{unsrt}

\newpage

\begin{figure} 
\includegraphics[width=3.5in]{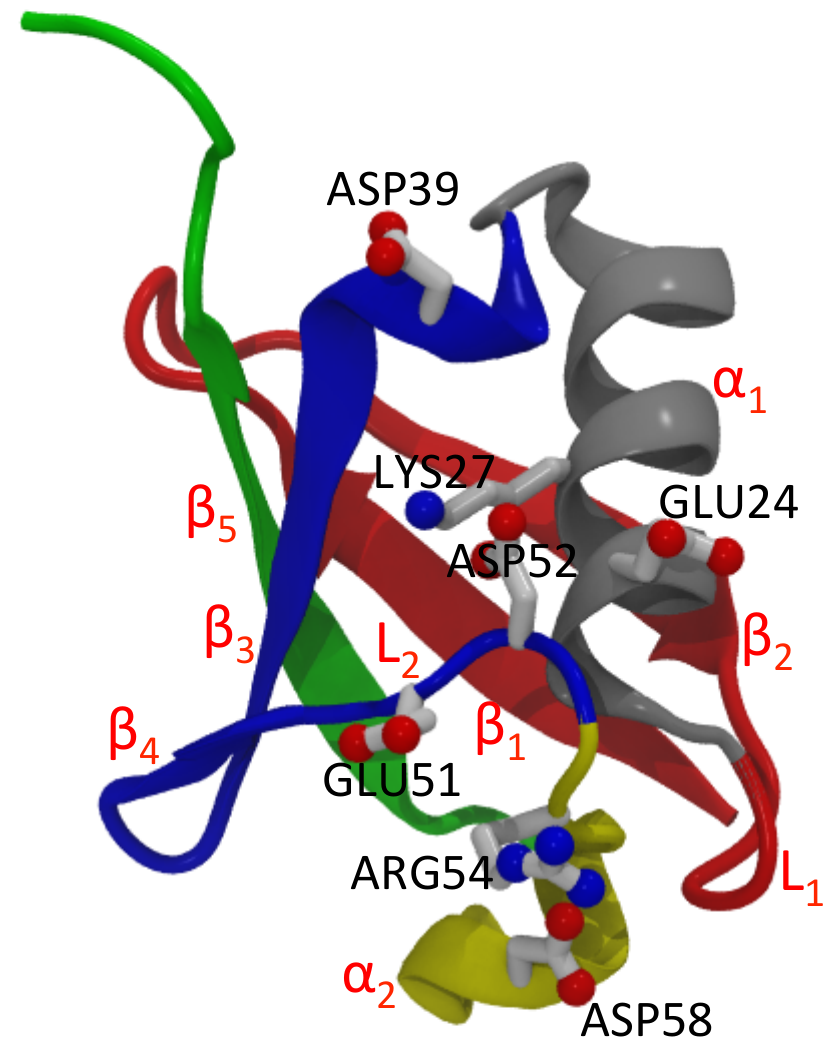}
\includegraphics[width=3.5in]{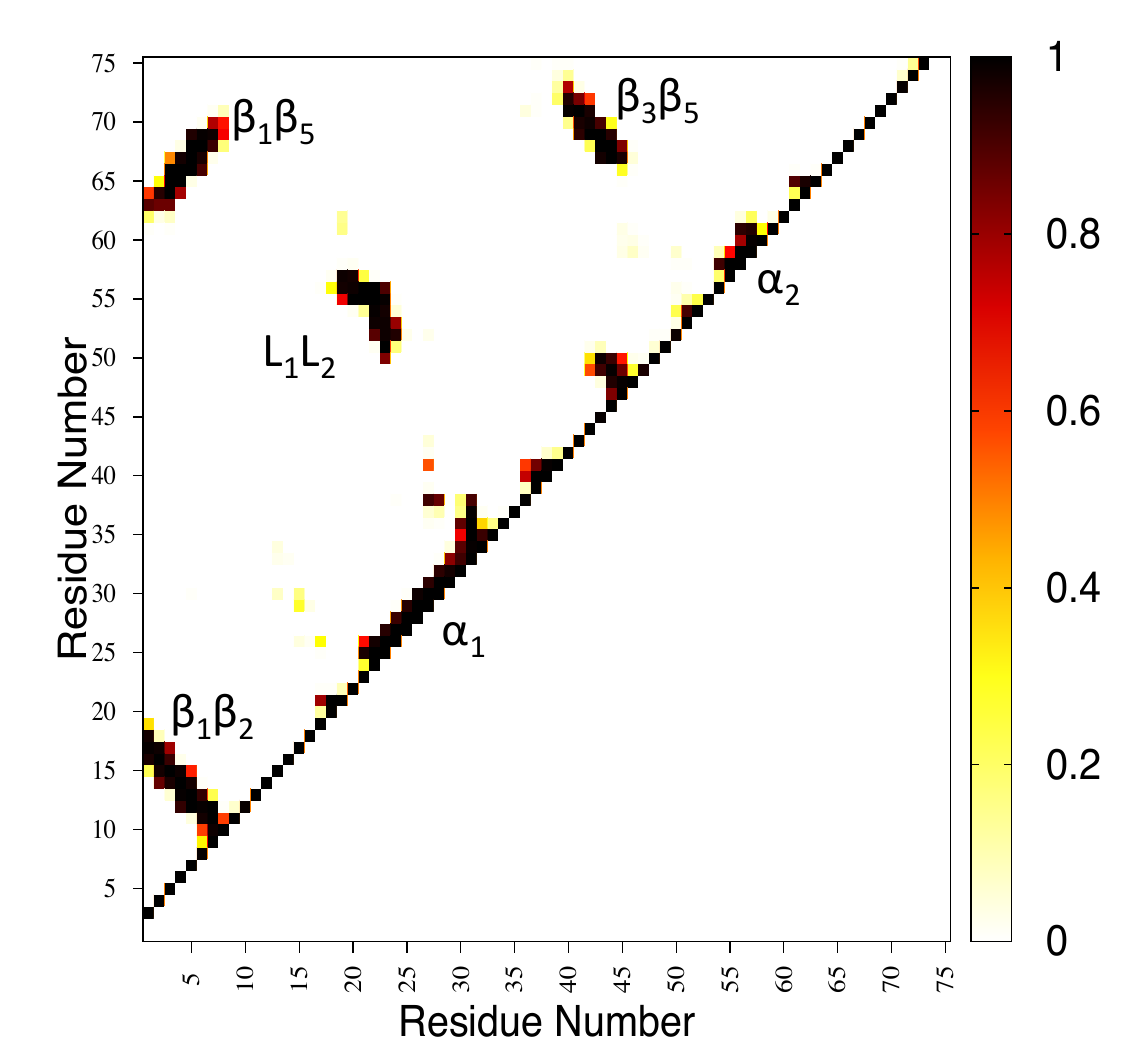}
\caption{(A) Crystal structure of Ub (PDB ID: 1UBQ). The charged residues, shown at the interface of the helix $\alpha_1$ and loops $L_1$ and $L_2$, stabilize the intermediate in the folding pathways of Ub in neutral pH.  (B) The C$_\alpha$ contact map of Ub shows contacts between various secondary structural elements in the folded state.}\label{figs0}
\end{figure}

\begin{figure}
\includegraphics[width=6.0in]{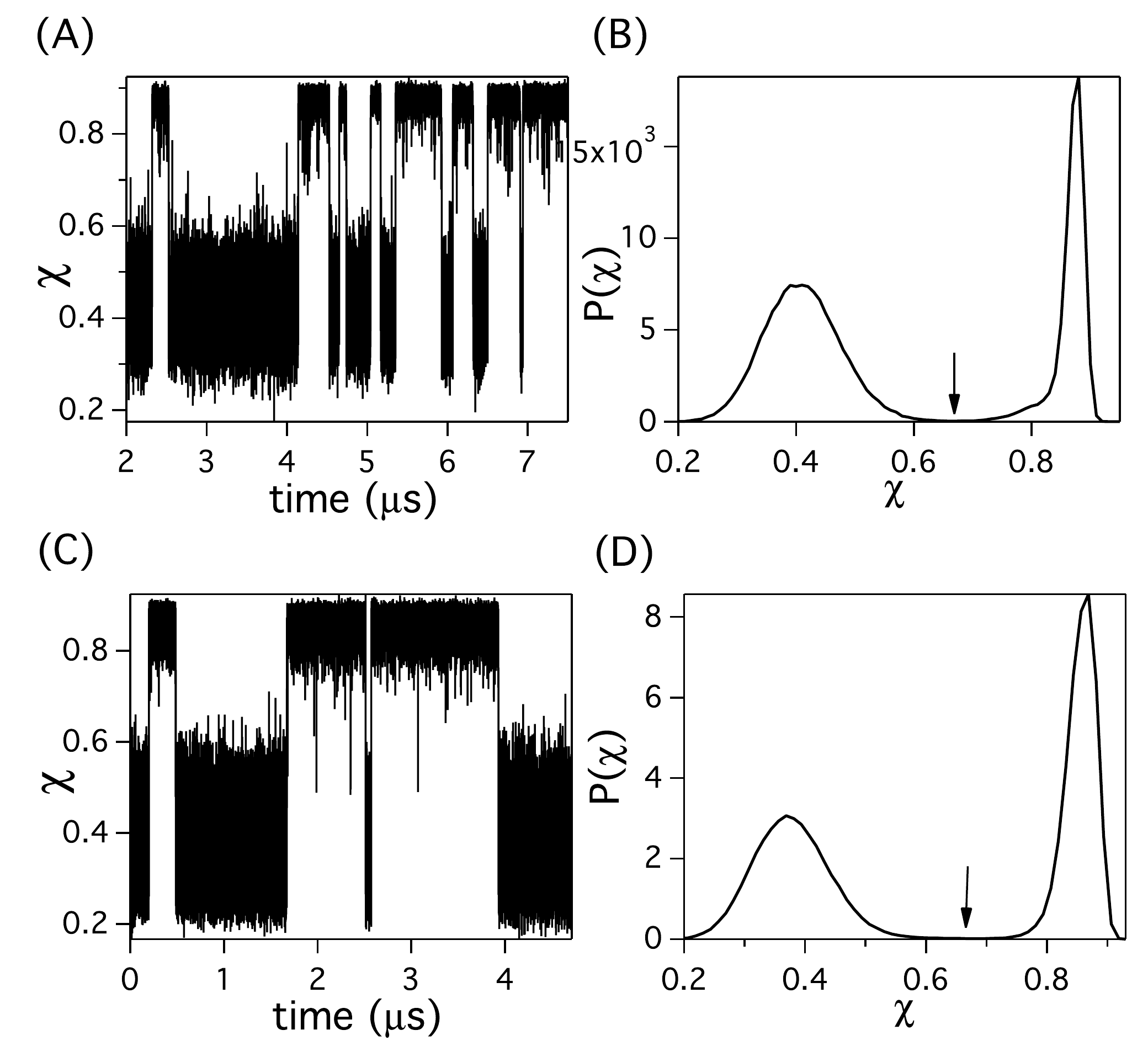}
\caption{(A) The structural overlap function\cite{Camacho93PNAS}, $\chi$, as a function of time for Ub at low pH at $T_m = 353K$. Ub undergoes multiple transitions between the UBA and NBA basins. (B) The probability distribution of $\chi$, $P(\chi)$ at $T_m$. The arrow is the value of $\chi = 0.67$ separating the UBA and NBA basins. (C) $\chi$ as a function of time for Ub at neutral pH at $T_m \approx 355K$. (D) The probability distribution of $\chi$, $P(\chi)$ at $T_m$ in neutral pH. The arrow at $\chi = 0.68$ separates the UBA and NBA.}\label{chi}
\end{figure}

\newpage

\begin{figure}
\includegraphics[width=5.0in]{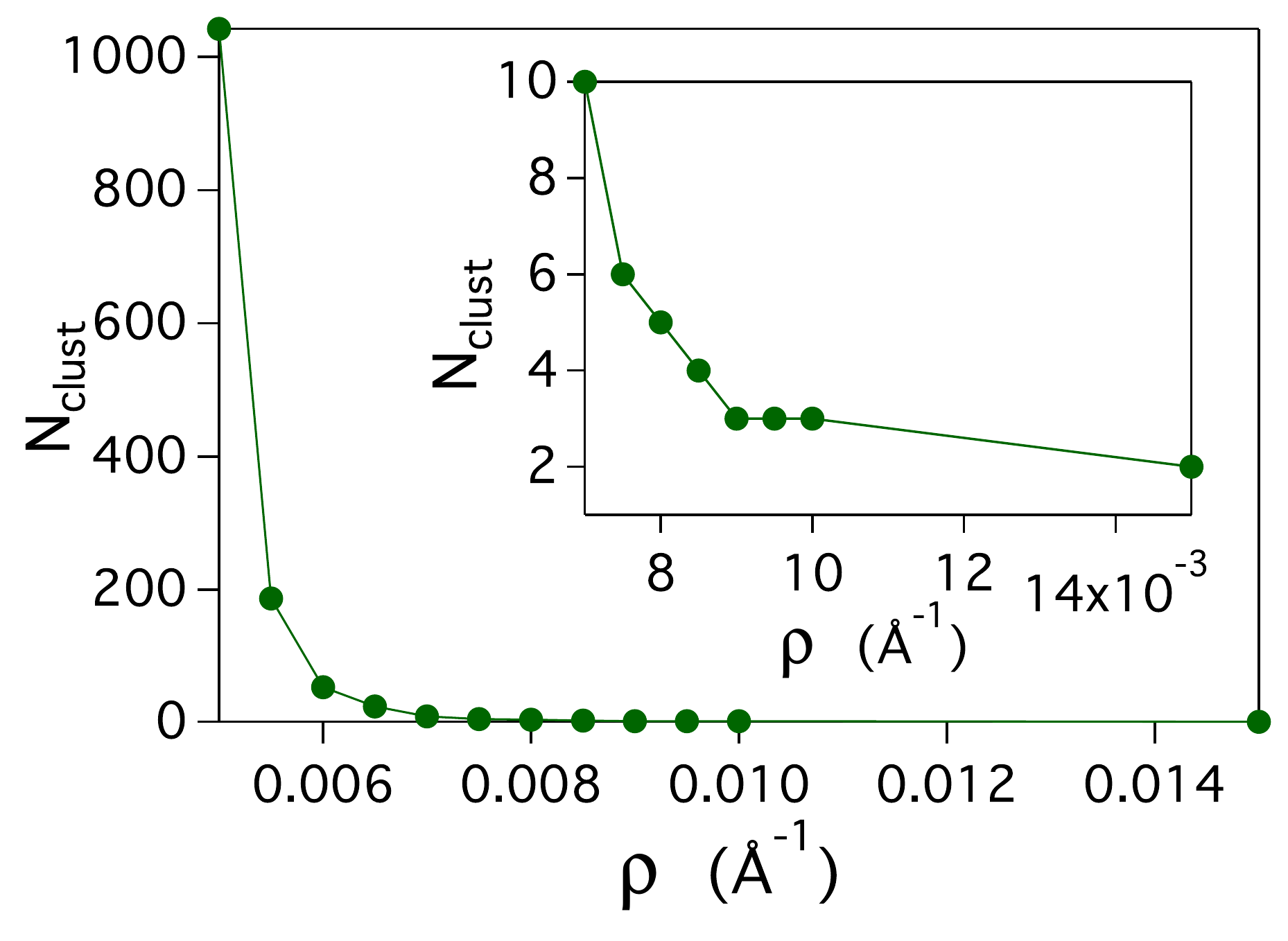} 
\caption{The number of clusters $N_{clust}$, in the Ub folding trajectory (shown in Fig.~S2A) at low pH and $T_m=353K$  as a function of the cutoff for the geometric distance $\rho$ (defined in the main text) between the conformations. The number of conformations increase exponentially as $\rho$ is decreased.}
\end{figure}

\newpage

\begin{figure}
\includegraphics[width=5.0in]{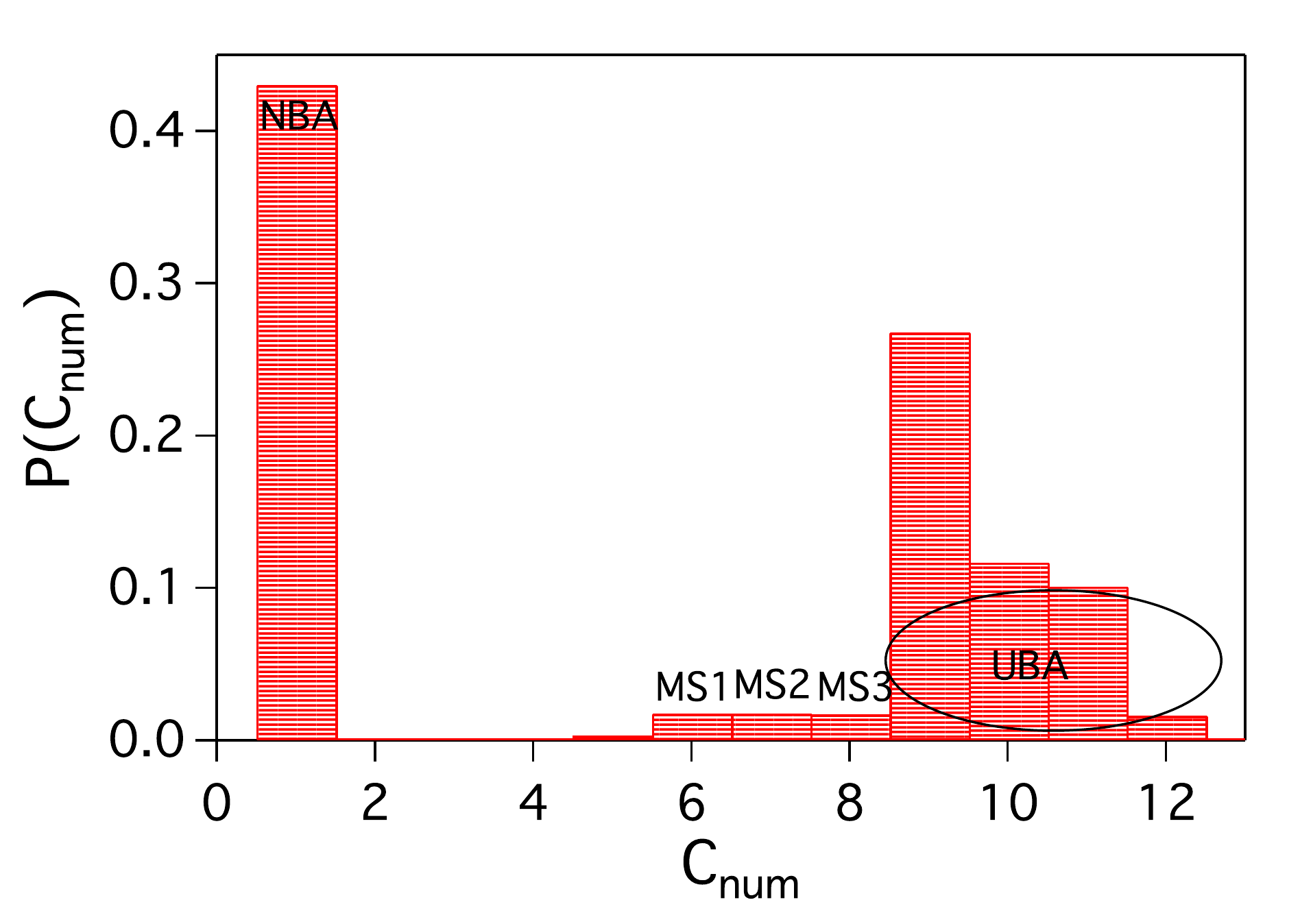} 
\caption{The probability distribution of the clusters identified in the protein folding trajectory for Ub at low pH and $T_m=353K$ (Fig.~S2A) using $\rho=0.055$ \AA$^{-1}$. Out of the 187 clusters identified, the 8 clusters (1, 6-12) have probabilities greater than 0.01. The cumulative probability of these 8 clusters is greater than 0.98. The 8 clusters based on the secondary structural features are further coarse-grained into 5 clusters, NBA (1), MS1 (6), MS2 (7), MS3(8), UBA(9-12).}
\end{figure}

\newpage

\begin{figure}
\includegraphics[width=3.5in]{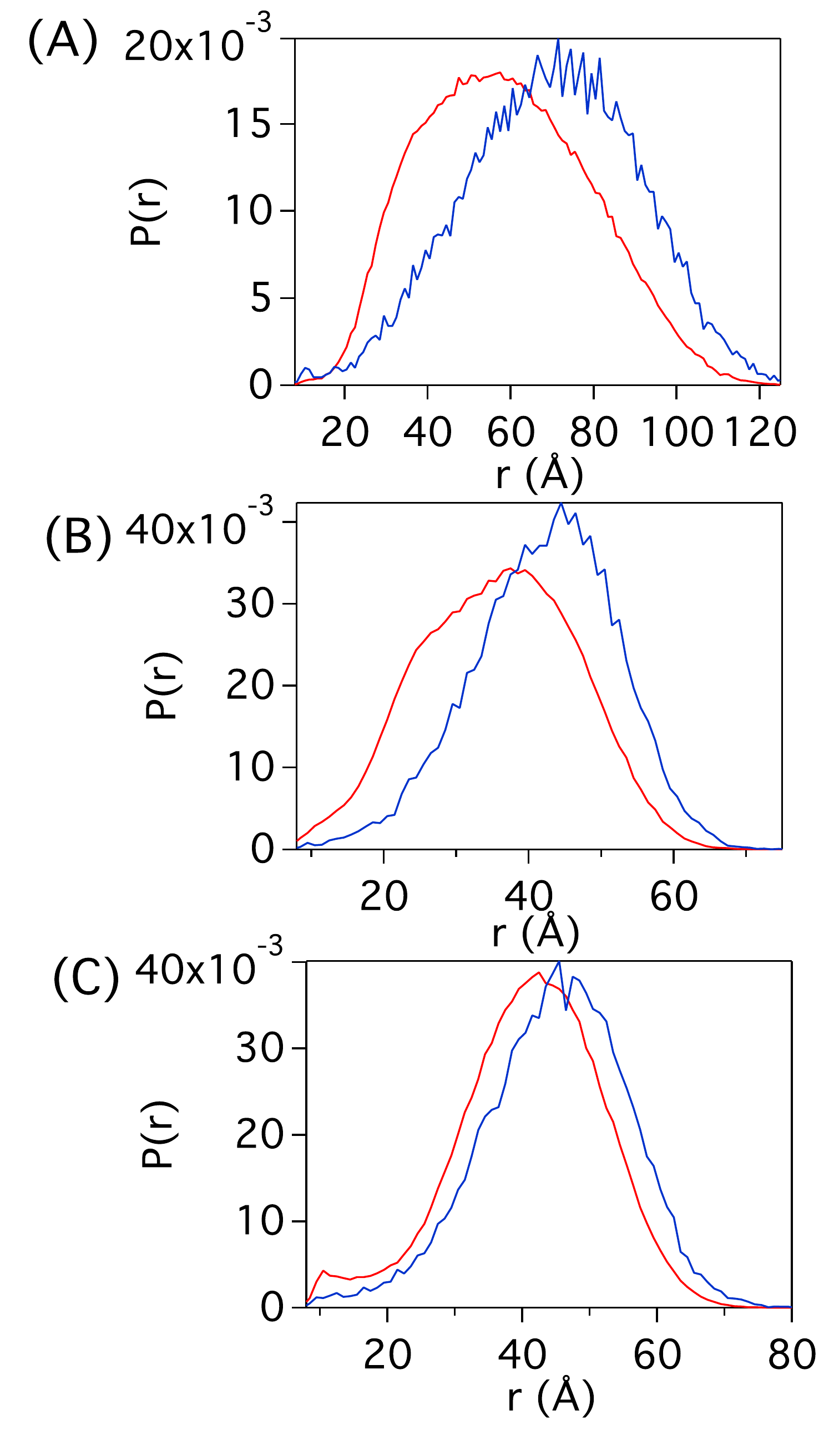} 
\caption{Probability distribution of the center of mass distance, $P(r)$, between the secondary structural elements at $T=400$K. (A) $\beta_1$ (residues 1 to 9) and $\beta_5$ (residues 66 to 72) (B) $\beta_3$ (residues 41 to 46) and $\beta_5$ (residues 66 to 72), (C) $\alpha_1$ (residues 20 to 27) and $L_2$ (residues 50 to 58). The data in red and blue represent neutral and low pH, respectively.}\label{figs4}
\end{figure}

\begin{figure}
\includegraphics[width=4.5in]{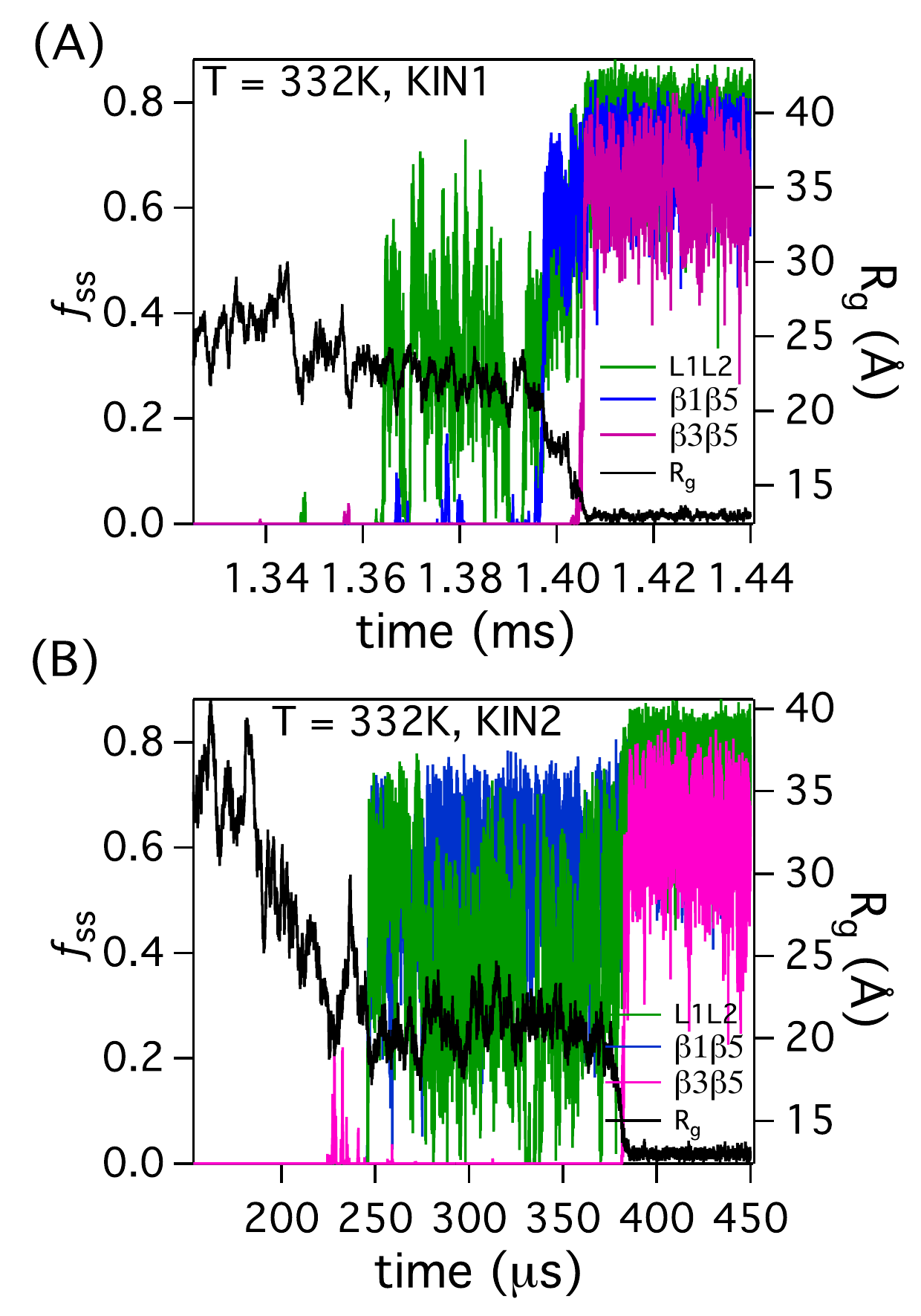}
\caption{Fraction of native contacts in various secondary structural elements ($\beta_1 \beta_5$, $\beta_3 \beta_5$, and $L_1L_2$), $f_{ss}$ and $R_g$ as a function of time in low pH trajectories folding at $T=332K$. The plots in the two panels are for trajectories labeled KIN1 and KIN2 in Fig.~6B. Green, blue and magenta colors correspond to secondary structures $L_1L_2$, $\beta_1 \beta_5$, and $\beta_3 \beta_5$ respectively. $R_g$ as function of time is in black}
\label{kin_T_332K}
\end{figure}

\newpage

\begin{figure}
\includegraphics[width=6.5in]{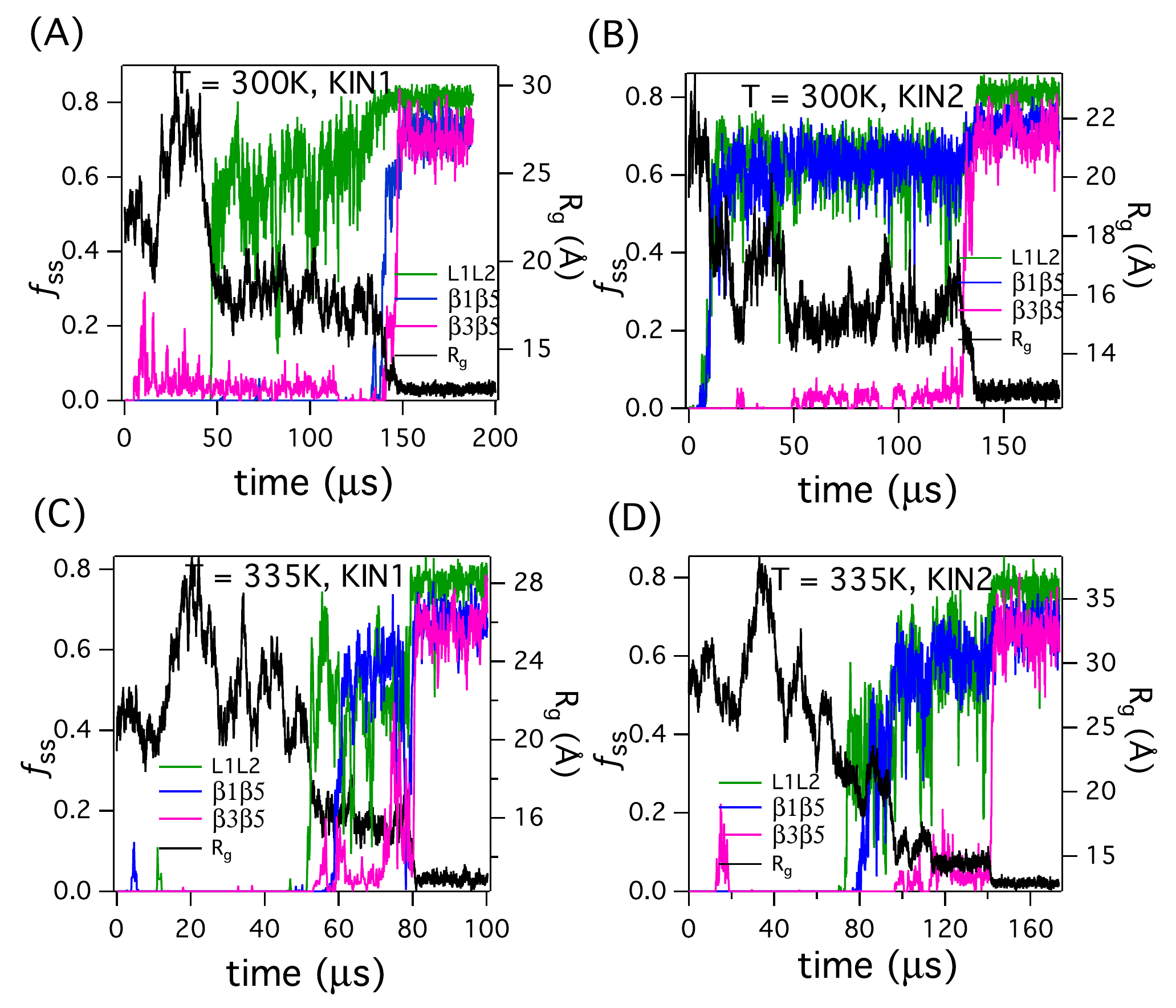}
\caption{Fraction of native contacts in various secondary structural elements ($\beta_1 \beta_5$, $\beta_3 \beta_5$, and $L_1L_2$), $f_{ss}$ and $R_g$ as a function of time in neutral pH. The plots are for trajectories shown in Fig.~8. (A) $T=300$K, KIN1 (B) $T=300$K, KIN2 (C) $T=335$K, KIN1 (D) $T=335$K, KIN2 . }
\label{kin_ele}
\end{figure}

\newpage
\begin{figure}
\includegraphics[width=5in]{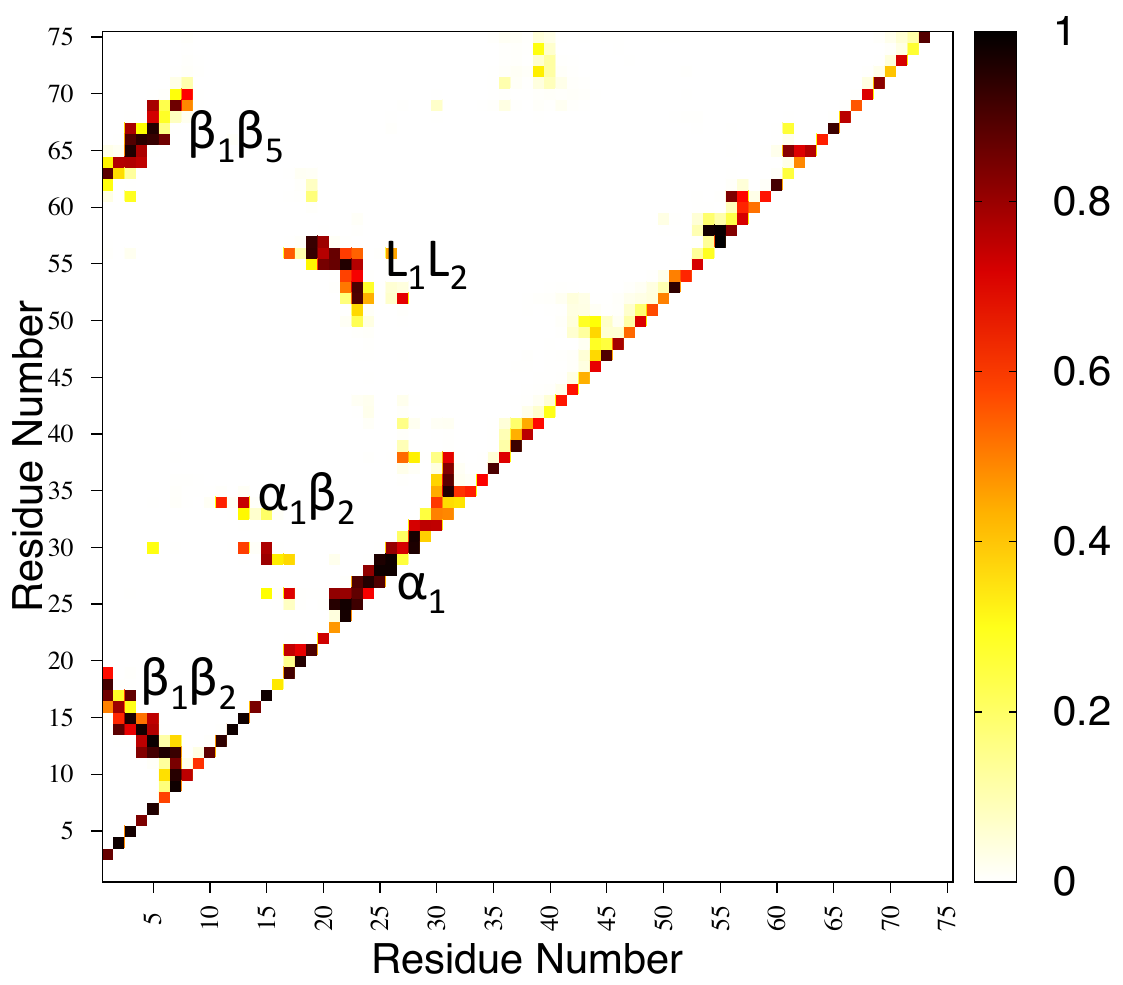} 
\caption{$C_\alpha$ contact map of the intermediate I3 (Fig.~8) observed in the Ub folding trajectories at neutral pH and $T=335K$ (Fig.~8). The contact map shows the contacts between the helix $\alpha_1$ and sheet $\beta_2$.}\label{figs6}
\end{figure}
\newpage

\begin{figure}
\includegraphics[width=5in]{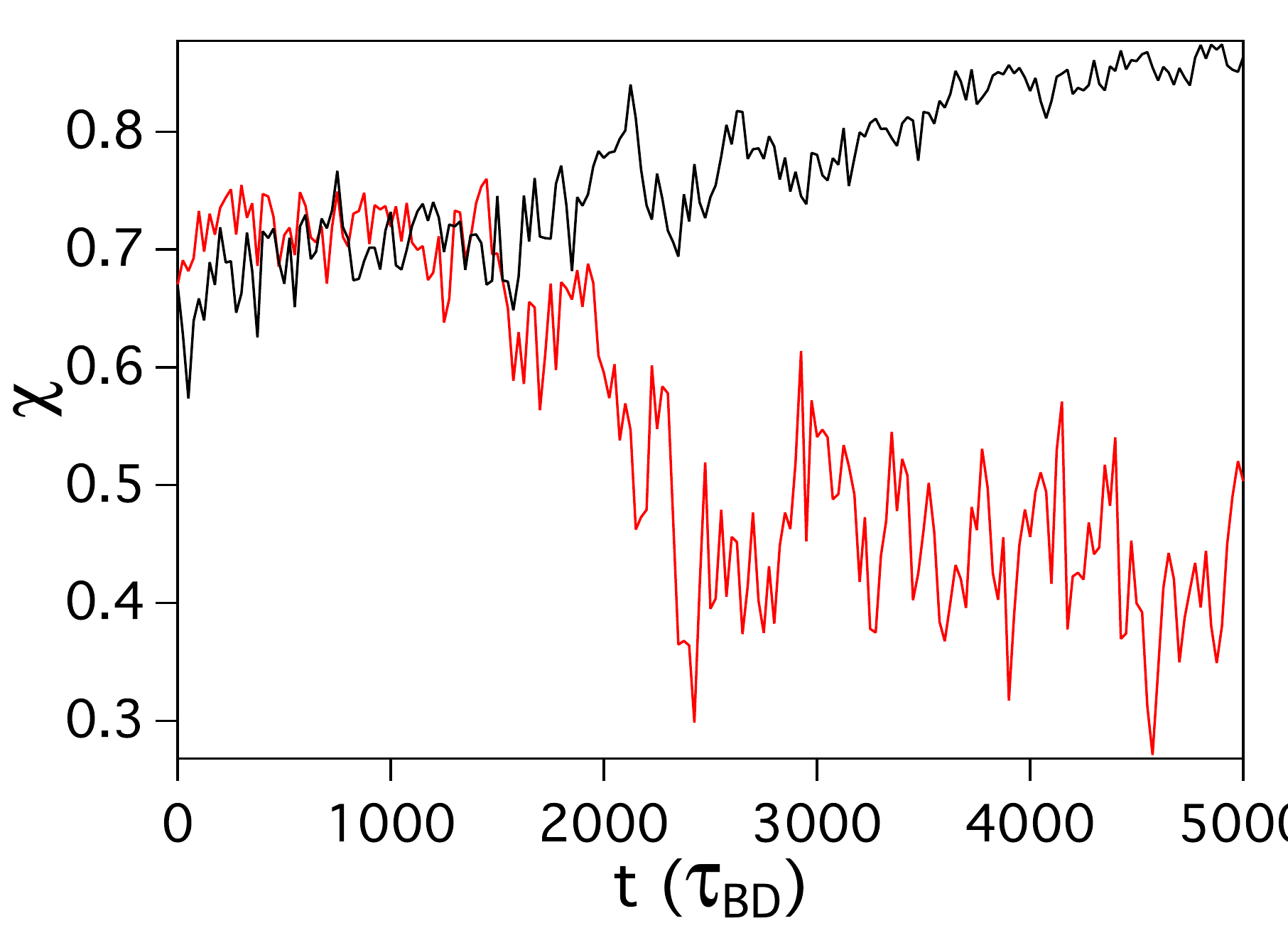} 
\caption{The structural overlap function, $\chi$ as a function of time, $t$, for two different trajectories spawned from one of the transition state structures of Ub at low pH with $T_m = 353K$ (Fig.~9).  Trajectory in red and black end up in the NBA and UBA respectively.}\label{figs5}
\end{figure}

\newpage

\begin{table}[ht]
\caption{Parameters for the SOP-Side Chain model}
\label{par}
\begin{center}
\begin{tabular}{|c | c|}
\hline
\hline
 Parameters & Protein \\
\hline
\hline
$R_o$ & 2.0 \\
$k$ & 20 kcal/(mol. \AA$^2$) \\
$R_c$ & 8 \AA  \\
$\epsilon_h^{bb}$~\footnote{Values are chosen such that the protein melting temperature in simulations is approximately in agreement with the experiments\cite{Wintrode94Proteins,Molero99Biochemistry}} & 0.45  kcal/mole  \\
$\epsilon_h^{bs  \  \ a}$& 0.45  kcal/mole  \\
$\epsilon_l^{ \ a}$ & 1.0  kcal/mole  \\
$\sigma^{bb}$ & 3.8 \AA  \\
$\epsilon$ & 10.0  \\
\hline
\hline
\end{tabular}
\end{center}
\end{table}

\newpage
 
\begin{table}[h]
\caption{Sidechain and backbone radii of amino acids based on partial molar volumes}
\label{radii}
\begin{center}
\begin{tabular}{ccc}
 Residue   & ~~~~~~~~~~~~~   &~~~~~~~~~Radius ($\AA$)~~~~~~~~~ \\
\hline \hline
$Gly$         &    &    $0$   \\
$Ala$         &    &    $2.52$   \\
$Val$         &    &    $2.93$   \\
$Leu$         &    &    $3.09$   \\
$Ile$         &    &    $3.09$   \\
$Met$         &    &    $3.09$   \\
$Phe$         &    &    $3.18$   \\
$Pro$         &    &    $2.78$   \\
$Ser$         &    &    $2.59$   \\
$Thr$         &    &    $2.81$   \\
$Asn$         &    &    $2.84$   \\
$Gln$         &    &    $3.01$   \\
$Tyr$         &    &    $3.23$   \\
$Trp$         &    &    $3.39$   \\
$Asp$         &    &    $2.79$   \\
$Glu$         &    &    $2.96$   \\
$Hse^a$       &    &    $3.04$   \\
$Hsd$         &    &    $3.04$   \\
$Lys$         &    &    $3.18$   \\
$Arg$         &    &    $3.28$   \\
$Cys$         &    &    $2.74$   \\
$backbone$ &    &    $2.25$   \\
\hline
\end{tabular}
\end{center}
$^aHse$ - Neutral histidine, proton on NE2 atom. $Hsd$ - Neutral
histidine, proton on ND1 atom.
\end{table}

\newpage